\documentclass[preprint,11pt,preprintnumbers,nofootinbib]{revtex4}
\usepackage{graphicx}% Include figure files
\usepackage{dcolumn}% Align table columns on decimal point
\usepackage{bm}% bold math
\usepackage{amsfonts}
\begin{document}

\date{\today}
%\title{Staggered Multi-Field Inflation: Gravity Waves and a New Class of Models}
\title{Multi-Field Inflation on the Landscape}

\author{Diana Battefeld} 
\email[email:]{diana.battefeld(AT)helsinki.fi}
\affiliation{Helsinki Institute of Physics, P.O. Box 64, University of Helsinki, FIN-00014 Helsinki, Finland\\ and APC, UMR 7164, 10 rue Alice Domon et Leonie Duquet, 75205 Paris Cedex 13, France}
\author{Thorsten Battefeld}
\email{tbattefe(AT)princeton.edu}
\affiliation{ Princeton University,
Department of Physics,
NJ 08544
}

\pacs{}
\begin{abstract}
We examine a wide class of multi-field inflationary models based on fields that decay or stabilize during inflation in a staggered fashion. The fields driving assisted inflation are on flat, short stretches, before they encounter a sharp drop; whenever a field encounters such a drop due to its slow roll evolution, its energy is transferred to other degrees of freedom, i.e. radiation. The rate at which fields decay is determined dynamically and it is not a free parameter in this class of models. To compute observables, we generalize the analytic framework of staggered inflation, allowing for more general initial conditions and varying potentials.  By searching for generic situations arising on the landscape, we arrive at a setup involving linear or hilltop potentials and evenly spread out initial field values. This scenario is not more fine tuned than large-field models, despite the fact that many more degrees of freedom are involved. Further, the $\eta$-problem can be alleviated.

The additional decrease of the potential energy caused by the decay of fields provides leading order contribution to observables, such as the scalar and tensor spectral index or the tensor to scalar ratio, for which we derive general expressions. We compare the predictions with WMAP5 constraints and find that hilltop potentials are borderline ruled out at the $2\sigma$-level, while linear potentials are in excellent agreement with observations. We further comment on additional sources of gravitational waves and non-Gaussianities that could serve as a smoking gun for staggered inflation.
\end{abstract}
\maketitle
\newpage

\tableofcontents

%%%%%%%%%%%%%%%%%%%%%%%%%%%%%%%%%%%%
\section{Introduction}
The embedding of single-field inflation within string theory is an active field of study, see \cite{McAllister:2007bg,Cline:2006hu,Burgess:2007pz,Kallosh:2007ig} for recent reviews. One of the most successful setups yet is the KKLMMT proposal \cite{Kachru:2003sx}. A complete implementation is not easily achieved since many conceptual challenges need to be overcome. Among these challenges is the requirement that all moduli fields need to be stabilized, that the inflaton potential ought to be extremely flat and that corrections to the potential must be understood well enough in order to guarantee  that the flatness is not spoiled once a field traverses a super-Planckian distance. Remarakably, it seems possible to addresses these challenges within the KKLMMT proposal, but at the cost of fine-tuning and a construction requiring delicately arranged ingredients \cite{Baumann:2008kq,Baumann:2007ah} (see however \cite{Hoi:2008gc}).

But if we turn our attention to a generic location on the landscape, single-field inflation appears unlikely:
moduli-fields can be dynamical, potentials are generically too steep and they change significantly if the fields roll far enough.  A way to circumvent some of these problems is to implement assisted inflation \cite{Liddle:1998jc,Malik:1998gy,Kanti:1999vt,Kanti:1999ie,Calcagni:2007sb}, a type of multi-field inflation, whereby many fields are dynamical and assist each other in driving the inflationary phase. In this scenario, the potentials can be steeper without spoiling slow roll, because a given field sees the driving force of its own potential, but it is slowed down by the combined Hubble friction of all fields. In addition, if the number of fields is large, none of them needs to traverse a super-Planckian stretch in fields space, thus alleviating the $\eta$-problem to some extent \cite{Liddle:1998jc,Kanti:1999vt,Kanti:1999ie}. From this point of view, multi-field models are more desirable and natural than single-field ones. Some recent realizations of assisted inflation are $\mathcal{N}$-flation \cite{Dimopoulos:2005ac,Easther:2005zr} (fields are identified with axions arising from some KKLT
compactification \cite{Kachru:2003aw} of type IIB string theory, see also \cite{Misra:2007cq,Misra:2008tx}), M5-brane inflation \cite{Becker:2005sg} (inflatons are identified with distances between branes that are spread out over an orbifold) or inflation from tachyons \cite{Piao:2002vf,Majumdar:2003kd}.

Of course, all moduli-fields need to be stabilized once inflation is over, that is moduli stabilization has to occur during or towards the end of assisted inflation.  Current models of multi-field inflation are often described by an effective single-field model, see i.e. \cite{Wands:2007bd} for a review; it is further assumed that all fields stabilize coherently, leading to a sudden end of inflation and (p)reheating just as in the single-field case \footnote{Note that reheating predominantly hidden sectors can be a problem for multi-filed models \cite{Green:2007gs}.}. However, the use of an effective single field is problematic for several reasons; first of all, there is no a priori reason to expect that fields decay (stabilize) all at once; indeed, there are cases, such as in inflation from multiple M5-Branes \cite{Becker:2005sg,Krause:2007jr,Ashoorioon:2006wc,Ashoorioon:2008qr}, where a coherent decay (stabilization) is impossible 
\footnote{During inflation the M5-branes separate from each other until they encounter a boundary brane into which they dissolve, one after the other.}. A proper computational framework has to incorporate this feature, such as in \cite{Battefeld:2008py,Battefeld:2008ur} where analytic tools for staggered inflation were developed (see also the numerical treatment in \cite{Ashoorioon:2006wc}, where this effect was named cascade inflation). Another problem of an effective single-field description is its inapplicability during reheating: even if inflation ends coherently for all fields, they generically de-phase during reheating, resulting in the suppression of many resonances \cite{Battefeld:2008bu,Battefeld:2008rd}, albeit tachyonic preheating \cite{Dufaux:2006ee} might still work \cite{Battefeld:2008rd,Braden,inprep2}. Further, the production of isocurvature perturbations and non-Gaussianities cannot be computed correctly.

To incorporate that fields decay or stabilize during inflation, we extend the formalism of \cite{Battefeld:2008py}, employing again the coarse grained description proposed therein: by smoothing out the number of fields $\mathcal{N}(t)$ we can introduce a continuous decay rate $\Gamma(t)\equiv -\dot{\mathcal{N}}/\mathcal{N}$. This rate causes energy transfer from the inflaton sector to an additional component of the energy momentum tensor $T^{\mu\nu}_r$, i.e. radiation. To avoid spoiling inflation, the ratio $\bar{\varepsilon}=3(1+w_r)\rho_r/(2\rho_{total})\simeq \Gamma/(2H)$ needs to be small. The resulting setup shares some similarities to warm inflation \cite{Berera:1995ie,Berera:2008ar} or dissipative inflation \cite{Hall:2007qw}, without many of its shortcomings (see i.e. \cite{Yokoyama:1998ju} for some problems within warm inflation related to the additional frition term, which we avoid). A study of scalar perturbations within staggered inflation reveals that $\bar{\varepsilon}$ appears alongside the usual slow roll parameters $\varepsilon$ and $\eta$ in observables \cite{Battefeld:2008py}. 

In \cite{Battefeld:2008py} several simplifying assumptions were made, such as identical initial field values and potentials for all inflatons. In this paper, we relax these conditions to allow for more general cases, but we still assume flat potentials. To be precise,  we assume that the change of the potential energy in a given field over its slow roll phase is small compared to its potential energy. Within this setup, we compute the spectrum of gravitational waves $\mathcal{P}_T$, the spectral index $n_{T}$ and the tensor to scalar ratio $r$. 

Equipped with this improved formalism, we construct new models of multi-field inflation motivated by generic potentials on the landscape \cite{Susskind:2003kw}, with $\mathcal{N}\sim 10^{3}$ fields and $\bar{\varepsilon}\gg \varepsilon$ \footnote{ $\varepsilon\sim \tilde{\varepsilon}\bar{\varepsilon}$ where $\tilde{\varepsilon}=\Delta V_i/V_i$; thus $\bar{\varepsilon}\gg \varepsilon$ is identical with our assumption of flat potentials.}: we arrive at models where fields can participate in assisted inflation until they encounter a sharp drop, at which point they decay or stabilize quickly. Thus, the decay rate is determined by the dynamics and it is not a free parameter. We assume similar potentials (hilltop or linear) for all fields, but allow for a narrow spread of masses or slopes. These types of potentials are motivated by expanding around either flat stretches or maxima on the landscape. Regarding initial conditions, we spread the fields evenly over the allowed interval and argue that this choice is the least fine tuned one. 

The scenario is similar in spirit to chain inflation \cite{Freese:2004vs,Feldstein:2006hm,Freese:2006fk,Huang:2007ek,Chialva:2008zw,Ashoorioon:2008pj,Ashoorioon:2008nh}, which is supposed to operate via successive, rapid tunnelling between different vacua on the landscape. However, instead of a rapid succession of first order phase transitions, which is difficult to achieve and can cause fatal problems in chain inflation related to bubble nucleation \cite{Ashoorioon:2008pj}, fields follow a smooth path through the landscape. The presence of such a path becomes more probable the higher the dimensionality of the moduli space is, but the danger of getting trapped in a meta-stable minimum remains.

Within staggered inflation, we compute observables, such as $n_s$ and $r$, that are dominated by $\bar{\varepsilon}$; a comparison with the WMAP5 data \cite{Komatsu:2008hk} reveals that hilltop potentials are already ruled out at the $2\sigma$-level, while linear potentials are in excellent agreement with observations. We also re-examine briefly inflation driven by tachyons, as proposed in \cite{Majumdar:2003kd}, and conclude that a constant decay rate is already ruled out in this setup while a serial condensation is still viable.   

We further comment on expected unique signatures of staggered inflation, such as additional gravitational waves and non-Gaussianities, that should arise whenever a field decays or stabilizes. To compute these signals, one has to go beyond the analytic formalism of this paper and study the decay of fields in detail. This would also provide a numerical check of the analytic framework. 

We conclude that staggered inflation offers a novel, compelling possibility to implement inflation within string theory, enabling the use of potentials that would be considered unsuitable otherwise. 

The concrete outline of this paper is as follows: we start by reviewing the background evolution of staggered inflaton in Sec.~\ref{sec:bgr}, which is used in our discussion of scalar and tensor perturbations in Sec. \ref{scalarperturbations} and \ref{sec:tensor}. We follow with an examination of concrete models, linear potentials in Sec.~\ref{sec:linear} and hilltop potentials in Sec.~\ref{sec:hilltop}, allowing for different slopes and masses in Sec.~\ref{sec:linearv} and \ref{sec:quadraticv}. The predictions are compared with WMAP5 constraints in Sec.~ \ref{sec:discussion}, where we also re-examine inflation driven by multiple tachyons. We elaborate on the theoretical motivation for these models in Sec.~\ref{sec:landscape} and point out open questions within staggered inflation in Sec.~\ref{sec:opensissues}, before concluding in Sec.~\ref{sec:conclusion}.

\section{Background Evolution \label{sec:bgr}}
\begin{figure}[tb]
\includegraphics[scale=0.6,angle=0]{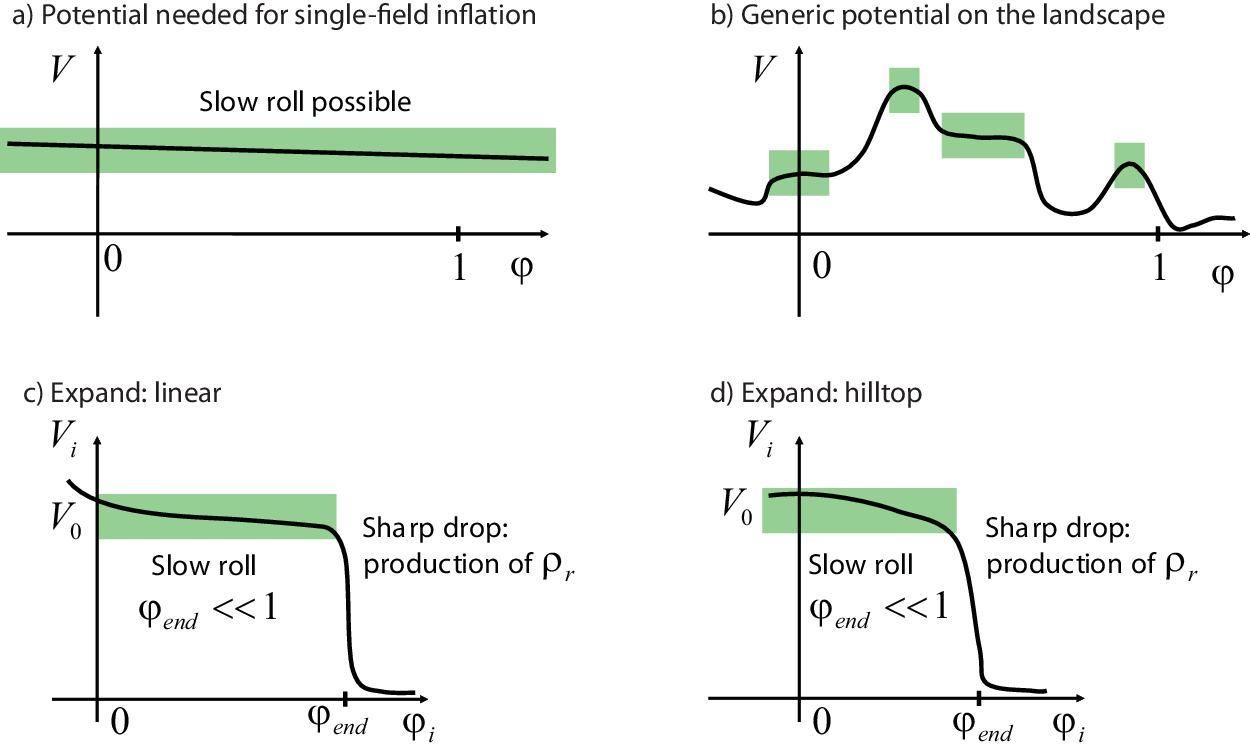}
   \caption{\label{pic:potential} Schematics of potentials, shaded regions allow for slow-roll inflation: a) Potential needed for single-field inflation (such as the KKLMMT proposal) that is flat for $\Delta \varphi \sim \mathcal{O}(10)$. b) A more generic potential on the landscape, with valleys, hills, and steep as well as some shallow regions. c) A region that is shallow until $\varphi_{end}\ll 1$ at which point the potential drops. If a field encounters this cliff, it decays (stabilizes) and $\rho_r$ is produced. We expand around this flat stretch, $V_i\approx V_0-c_i\varphi_i$ . d) A maximum, around which we also expand, $V_i\approx V_0-m_i^2\varphi_i^2/2$.}
\end{figure}

Consider $\mathcal{N}\sim \mathcal{O}(10^3)$ uncoupled inflaton fields $\varphi_i$, $i=1\dots \mathcal{N}$, with potentials
\begin{eqnarray}
V_i=V_0-f_i(\varphi_i)\,, \label{potential1}
\end{eqnarray}
where $f_i(\varphi_i)$ is a monotonically decreasing function, and the total potential is $W\equiv \sum_i V_i$ \footnote{$W$ should not be confused with the super-potential in SUGRA.}. We focus on those fields with a large $V_0$ contributing the most to inflation, and ignore fields with much smaller $V_0$. Therefore, we set $V_0$ identical for the fields we consider. Regarding the free functions $f_i(\varphi_i)$, we are primarily interested in two cases: first, we identify inflatons with fields on the landscape which, by chance, encounter a flat region in the potential; we can then expand around such a position and get linear potentials $f_i=c_i\varphi_i$. If, on the other hand, the fields are close to a maximum, we expand again and get quadratic potentials $f_i=m_i^2\varphi_i^2/2$ (see Fig.~\ref{pic:potential} for a schematic). The latter case appears less likely given a generic position on the landscape, but since no canonical measure is known (see Sec.\ref{sec:landscape} for more details), we discuss both cases. 

In either setup, only narrow distributions of $c_i$ or $m_i$ are of interest to us, because in the presence of wildly different slopes, fields roll down quickly along the steeper ones and become irrelevant for inflation. For this reason we take $f_i= f_j\equiv f$ first, but relax this simplifying assumption as we proceed.

We expect an expansion around either a maximum or a shallow region to be valid only for a limited stretch in field space before the potential changes drastically. To model this we use (\ref{potential1}) only up until $\varphi_i=\varphi_{end}\ll 1$; once a field equals $\varphi_{end}$, we assume that it encounters a sharp drop in the potential, falls down and converts its entire potential energy into some other form $\rho_r$, such as radiation.  Since we would like to model flat stretches we demand
\begin{eqnarray}
\tilde{\varepsilon}\equiv \frac{f(\varphi_{end})}{V_0}\ll 1\,. \label{tildevarepsilon}
\end{eqnarray}
In the following, we denote equality up to first order in small parameters with ``$\simeq$''.

If the fields start out from different initial values they encounter the drop in $V_i$ one after the other and decay in a staggered fashion. If there are many fields so that several fields decay in a given Hubble time, we can apply the analytic formalism of staggered inflation as proposed in \cite{Battefeld:2008py,Battefeld:2008ur}. We  need to amend this formalism slightly to accommodate more general initial conditions as well as varying slopes and masses (Sec. \ref{sec:linearv} and \ref{sec:quadraticv}).

Thus, let us briefly review (and extend) this formalism:  first, we promote the number of fields $\mathcal{N}$ to a smooth, time dependent function, so that we can introduce the continuous decay rate 
\begin{eqnarray}
\Gamma \equiv -\frac{\dot{\mathcal{N}}}{\mathcal{N}}>0\,. 
\end{eqnarray}
This rate is not a free parameter but follows once the initial values for the inflatons are specified and a solution to their slow roll equations of motion is found. If inflation is to last several e-folds, we need
\begin{eqnarray}
\varepsilon_{\mathcal{N}}\equiv \frac{\Gamma}{2H} \ll 1 \,,
\end{eqnarray}
otherwise the majority of fields encounter $\varphi_{end}$ in a single Hubble time. In the absence of a pre-inflationary model dictating the initial conditions, we distribute the fields evenly over the interval $\varphi_{ini}\dots \varphi_{end}$, namely
\begin{eqnarray}
\varphi_i(0)=\varphi_{ini}+\frac{\varphi_{end}-\varphi_{ini}}{\mathcal{N}(0)}(i-1)\,;\label{inicon}
\end{eqnarray}
this seems to be the most natural choice to us, but other initial distributions are possible (see Sec.\ref{sec:landscape} for heuristic arguments motivating this choice). Note that quantum mechanical fluctuations generically introduce $\delta \varphi_i^{QM}\sim H$. Hence, for hilltop potentials one should take $\varphi_{ini}\geq H$ in order for the classical evolution to dominate \footnote{The COBE normalization of the power-spectrum implies $\varphi_{ini}\geq 10^{-6}$ in the present study, see section \ref{scalarperturbations}.}.   

Given these initial conditions, the fields evolve according to their Klein-Gordon equations. During slow roll, that is if the slow roll parameters $\varepsilon_i\equiv (\partial V_i/\partial \varphi_i)^2/(2W^2)$, $\varepsilon\equiv\sum_i\varepsilon_i$ as well as $\eta_i\equiv (\partial^2 V_i/\partial \varphi_i^2)/W$ and $\eta\equiv \sum_i \eta_i$ are small, the equations of motion become
\begin{eqnarray}
3H\dot{\varphi}_i\simeq -\frac{\partial V_i}{\partial \varphi_i}\,. \label{slowroll}
\end{eqnarray}
Here, a dot denotes a derivative with respect to cosmic time and we set the reduced Planck mass equal to one, $m_{pl}^{-2}=8\pi G\equiv 1$. Eq.(\ref{slowroll}) has to be solved in conjunction with the Friedman equation
\begin{eqnarray}
3H^2=\rho_{total}\,,
\end{eqnarray}
where $\rho_{total}=\sum_i \rho_{\varphi_i}+\rho_r\equiv \rho_{I}+\rho_r$. The energy transfer from $\rho_I$ to $\rho_r$ manifests itself in the continuity equation, which becomes (see \cite{Battefeld:2008py} and also \cite{Watson:2006px} for related work where inflation is driven by a dynamically relaxing cosmological constant.)
\begin{eqnarray}
\dot{\rho}_I=-3H(\rho_I+p_I)+\dot{\mathcal{N}}V_0\left(1+\mathcal{O}(\tilde{\varepsilon})\right)\,. \label{cont1}
\end{eqnarray}
Since $\nabla_\mu T_{total}^{\mu 0}=0$ has to hold, we necessarily need the additional component of the energy momentum tensor to obey 
\begin{eqnarray}
\dot{\rho}_r=-3H(\rho_r+p_r)-\dot{\mathcal{N}}V_0\left(1+\mathcal{O}(\tilde{\varepsilon})\right)\,. \label{cont2}
\end{eqnarray}
In order for inflation to last several e-folds we need $\rho_r \ll \rho_{total}$, that is
\begin{eqnarray}
\bar{\varepsilon}\equiv \frac{3}{2}(1+w_r)\frac{\rho_r}{\rho_r+\rho_{I}} \ll 1\,,
\end{eqnarray}
where $w_r$ is the equation of state parameter of the additional component, $p_r=w_r\rho_r$. With these definitions, one can further simplify (\ref{cont1}) and (\ref{cont2}) to \cite{Battefeld:2008py}
\begin{eqnarray}
\dot{\rho}_I & \simeq & -2H(\varepsilon_{\mathcal{N}}+\varepsilon)\rho_I\,,\\
\dot{\rho}_r& \simeq & -2H \left(\frac{3}{2}(1+w_r)\rho_r -\varepsilon_{\mathcal{N}}\rho_{I}\right)\simeq 2H (\varepsilon_{\mathcal{N}}-\bar{\varepsilon})\rho_I\,,
\end{eqnarray}
where we ignored contributions that are second order in small parameters. As a result, $\rho_r$ approaches a scaling solution that tracks the inflationary energy $\rho_r\simeq \varepsilon_{\mathcal{N}}2\rho_{I}/(3+3w_r)$ for which
\begin{eqnarray}
 \varepsilon_{\mathcal{N}}\simeq \bar{\varepsilon}\,.
 \end{eqnarray}
 Further, the Hubble slow evolution parameter becomes \cite{Battefeld:2008py}
 \begin{eqnarray}
 \hat{\varepsilon}\equiv -\frac{\dot{H}}{H^2}\simeq \varepsilon+\bar{\varepsilon}\,.\label{Hubbleslow}
 \end{eqnarray}
 Note that $\bar{\varepsilon}$ appears alongside $\varepsilon$ and can be more important than the latter. As mentioned in \cite{Battefeld:2008py}, this type of staggered inflation is reminiscent to warm inflation \cite{Berera:1995ie}, without sharing many of its problems.
 
\section{Perturbations}
Scalar perturbations are discussed in \cite{Battefeld:2008py};  here we summarize the main results of \cite{Battefeld:2008py}, before proceeding to compute gravitational waves. Let's first briefly scrutinize the models of interest: we would like to focus on setups that possess a shallow potential, either because the fields are near a maximum, or because such fields have, by chance, encountered a shallow stretch. We parameterized this requirement by demanding that $\tilde{\varepsilon}$ from (\ref{tildevarepsilon}) is small, e.g. $\tilde{\varepsilon}\sim 10^{-2}$. For the models of interest the slow roll parameter becomes of order $\varepsilon \sim \bar{\varepsilon}\tilde{\varepsilon}\ll \bar{\varepsilon}$ \footnote{For instance, in the linear case $\varepsilon \simeq c^2/(2\mathcal{N}V_0^2)$, while $\bar{\varepsilon}\simeq \varepsilon_{\mathcal{N}}\simeq c/(2\varphi_{end}\mathcal{N}V_0)$ from (\ref{barepsilonlinear}) and $\tilde{\varepsilon}=c\varphi_{end}/V_0$, so that $\varepsilon\simeq \tilde{\varepsilon}\bar{\varepsilon}$. Similarily, using (\ref{barvarepsilonhilltop}) one can show that $\varepsilon \lesssim \tilde{\varepsilon}\bar{\varepsilon}/2$ for hilltop potentials.}. Hence, the contributions due to slow roll to observable parameters are negligible compared to those originating from the decaying fields, that is compared to those caused by $\varepsilon_{\mathcal{N}}\simeq \bar{\varepsilon}$ if $\tilde{\varepsilon}$ is small. Thus, we drop all terms proportional to $\varepsilon$.

\subsection{Scalar Perturbations \label{scalarperturbations}}
As argued in \cite{Battefeld:2008py}, isocurvature perturbations are suppressed in the present framework\footnote{There is a possibility that isocurvature perturbations are produced during the short intervals when fields decay; we assume that this does not occur, but this caveat clearly warrants further study.}, hence we focus on adiabatic perturbations; these are properly described by (the Fourier modes of) the Mukhanov variable $v_k=z\zeta_k$ with $z=1/\theta$, $\theta^2=1/(3a^2(1+w))$ and $w=p_{total}/\rho_{total}$ \cite{Mukhanov:1990me}. Here and in the following  we use $|c_s|\simeq 1$ ($c_s$ is  the adiabatic sound speed), since $\rho_r\propto \rho_I$ during inflation. If $\hat{\varepsilon}$ from (\ref{Hubbleslow}) evolves slowly, which is the case during inflation, we can approximate
\begin{eqnarray}
a(\tau)\propto (-\tau)^{-(1+\hat{\varepsilon})}\,, \label{bgra}
\end{eqnarray}
where $\tau=-\infty \dots 0$ is conformal time, $a\, d\tau=dt$ and $\partial (\,\,)/\partial \tau=(\,\,)^\prime$, solve the equation of motion for $v_k$
\begin{eqnarray}
v_k^{\prime\prime}+\left(k^2-\frac{z^{\prime\prime}}{z}\right)v_k=0
\end{eqnarray}
analytically in terms of Hankel functions \cite{Mukhanov:1990me,Bassett:2005xm}, match to the Bunch Davis vacuum in the far past ($v_k=\exp(-ik\tau)/\sqrt{2k}$), expand the solution on large scales and compute the curvature perturbation on uniform density surfaces $\zeta_k$. After some algebra, the resulting power spectrum for  $\zeta$ becomes, to leading order in $\bar{\varepsilon}$  (see \cite{Battefeld:2008py} for details)
\begin{eqnarray}
\mathcal{P}_{\zeta}&\equiv &\frac{k^3}{2\pi^2}|\zeta_k|^2\\
%&\simeq &\frac{G H^2}{\pi\bar{\varepsilon}}\label{powerscalar}\\\
&\simeq &\frac{1}{8\pi^2\bar{\varepsilon}}\frac{ H^2}{m_{pl}^2}\,,\label{powerscalar}
\end{eqnarray}
where we restored the reduced Planck mass \footnote{For comparison, the power spectrum in \cite{Battefeld:2008py}, where fields start out from identical initial values and the slow roll contribution is kept, reads $\mathcal{P}_{\zeta}
\simeq H^2/(8\pi^2 m_{pl}^2(\varepsilon \gamma^2 + \bar{\varepsilon}))$, where $\gamma$ is of order one and it is set by $\Gamma$ and the potential (see \cite{Battefeld:2008py} for details).}. Note that the amplitude of the power spectrum is set by the COBE normalization $P_{\zeta}=(2.41\pm 0.11) \times 10^{-9}$ \cite{Komatsu:2008hk}. With $\bar{\varepsilon}\sim 10^{-2}$, this implies an inflationary scale of $H\sim 10^{-6}$, which in turn gives $\varphi_{ini}\sim \delta \varphi_i^{QM}\sim H\sim 10^{-6}$ in hilltop potentials. Further, the scalar spectral index becomes \cite{Battefeld:2008py}
\begin{eqnarray}
n_s-1&=&\frac{d\,\ln \mathcal{P}_{\zeta}}{d\, \ln k}\\
&\simeq & (\delta-3)  \bar{\varepsilon}\,, \label{scalarspectralindex}
\end{eqnarray}
where 
\begin{eqnarray}
\delta \equiv \frac{\dot{\Gamma}H}{\Gamma \dot{H}}
\end{eqnarray}
measures the time dependence of the decay rate $\Gamma$, which also needs to be evaluated at around $N\approx 60$ e-folds before the end of inflation. Note that the condition for the validity of our approach (several fields ought to drop out in a given Hubble time for the smoothing of $\mathcal{N}(t)$ to be a good approximation) needs to be satisfied only when the Fourier modes $\zeta_k$ of interest leave the Hubble radius around $N\approx 60$; the reason being that once these modes leave the horizon they freeze out $\dot{\zeta}_k\rightarrow 0$.

\subsection{Tensor Perturbations \label{sec:tensor}}
Gravitational waves are less influenced by staggered inflation simply because they are not directly coupled to scalars. To see this, consider the amplitude $h$ of the two possible polarization states of a general tensor perturbation $h_{ij}=h(t)e_{ij}^{(+,\times)}(x)$, where $e_{ij}$ are eigen-modes of the spatial Laplacian \cite{Bassett:2005xm}. Fourier modes of this amplitude obey
\begin{eqnarray}
\ddot{h}_k+3H\dot{h}_k+\frac{k^2}{a^2}h_k=0\,;
\end{eqnarray}
hence, $h$ does not couple to the scalar sector, but is determined entirely by background quantities, once initial conditions are specified. We can eliminate the friction term by switching to conformal time and introducing $u_k\equiv ah_k/2$ so that
\begin{eqnarray}
u^{\prime\prime}_k+\left(k^2-\frac{a^{\prime\prime}}{a}\right)u_k=0\,.
\end{eqnarray}
Imposing quantum mechanical initial conditions in the far past $u_k=\exp(-ik\tau)/\sqrt{2k}$, just as for scalar perturbations, and using (\ref{bgra}), we find
\begin{eqnarray}
u_k=\frac{\sqrt{-\pi \tau}}{2}H^{(1)}_{\nu_\tau}(-k\tau)\,,
\end{eqnarray}
where $H^{(1)}_{\nu_\tau}$ is the Hankel function of the first kind with index $\nu_{\tau}=3/2+\hat{\varepsilon}$. On large scales, this can be expanded to 
\begin{eqnarray}
|u_k|=\frac{\sqrt{-\pi \tau}}{2\pi}\bar{\Gamma}(\nu_\tau)\left(-\frac{k\tau}{2}\right)^{-\nu_\tau}\,,
\end{eqnarray} 
where $\bar{\Gamma}$ is the Gamma-function. Thus, the amplitude of the tensor power spectrum becomes \cite{Watson:2004aq}
\begin{eqnarray}
\mathcal{P}_{T}&\equiv &2\frac{4\pi k^3}{(2\pi)^3}\left|h_k^2\right|\\
%&=&\frac{16}{\pi}GH^2 \label{powertensor}\\
&\simeq &\frac{2}{\pi^2}\frac{H^2}{m_{pl}^2}\,, \label{powertensor}
\end{eqnarray}
where we restored the reduced Planck mass in the last step and used $\bar{\Gamma}(3/2)=\sqrt{\pi}/2$ \footnote{The additional factor of $2$ in the power spectrum accounts for the two polarization modes of gravity waves.}. Further, the tensor spectral index reads
\begin{eqnarray}
n_{T}&\equiv& \frac{d\, \ln \mathcal{P}_T}{d\, \ln k}\\
&\simeq &-2\bar{\varepsilon} \,. \label{tensorspectralindex}
\end{eqnarray}
Computing the ratio of (\ref{powertensor}) to (\ref{powerscalar}) yields the tensor to scalar ratio
\begin{eqnarray}
r&\equiv &\frac{\mathcal{P}_T}{\mathcal{P}_\zeta}\\
 &\simeq &16 \bar{\varepsilon}\,. \label{tensortoscalarratio}
\end{eqnarray}
Note that we still have $r\simeq -8n_T$, just as for plain slow roll inflation, but both parameters are now determined by $\bar{\varepsilon}\simeq \varepsilon_{\mathcal{N}}=\Gamma/(2H)$ instead of $\varepsilon$.

\section{Models}
In string theory, many degrees of freedom (branes, fluxes, size and shape moduli etc.) are present in the early universe, all of which can be dynamical. However, in most approaches to inflation all but one or two degrees of freedom are already stabilized during inflation. The argument for this approach is simple: even if all degrees of freedom are dynamical at the onset of inflation, only those acted upon by very weak forces, that is those with a very flat potential, will remain dynamical for a long period of time, hopefully driving inflation, whereas all the other ones quickly relax to stable configurations. Thus, by focusing on only one or two degrees of freedom, one assumes that flat directions are so rare that one would be lucky to encounter one suitable field  for inflation.

However, this line of reasoning misses one crucial point: in the presence of many fields, each field sees the driving force of its own potential, but it is slowed down by the combined Hubble friction of all fields (this is the idea behind assisted inflation \cite{Liddle:1998jc}). It is then not far fetched to employ many fields which, by chance, find themselves on a flat stretch on the landscape and can drive inflation. Of course, once they encounter a sharp edge, which should happen for $\Delta \varphi_i \ll 1$, they rapidly fall down, convert some energy to other degrees of freedom such as radiation, and become irrelevant for inflation. For this type of inflation, one might expand around the flat stretch and use a linear approximation to the potential up until the cliff at $\varphi_{end} \ll 1$, as indicated in section \ref{sec:bgr}. In the next subsection, we discuss this model, followed by hilltop inflation. 

\subsection{Linear Potentials: $V_i=V_0-c\varphi_i$ \label{sec:linear}}
Consider $\mathcal{N}$ fields, with potentials $V_i=V_0-c\varphi_i$, that are valid up until $\varphi_{end}$, where we assume that the potential drops to zero. We distribute these fields according to (\ref{inicon}) with $\varphi_{ini}=0$, and apply the formalism of the previous sections to compute observables. To do so, we first need to extract $\Gamma$ (we determine $\Gamma$ without smoothing out the number of fields). Given the linear potential and a constant $\mathcal{N}\sim 10^3$  we can compute the evolution of the fields from (\ref{slowroll}) to
\begin{eqnarray}
\varphi_i(t)=\varphi_i^{ini}+\frac{c}{\sqrt{3\mathcal{N}V_0}}t\,,
\end{eqnarray}
(whenever a field decays, $\mathcal{N}$ decreases by one in this expression).
Thus, the time it takes for the field nearest to the edge to encounter the drop is  given by $\varphi_{\mathcal{N}}(\Delta t)=\varphi_{end}$ and reads to leading order in $1/\mathcal{N}$
\begin{eqnarray}
\Delta t\simeq \frac{\varphi_{end}}{c}\sqrt{\frac{3V_0}{\mathcal{N}}}\,.
\end{eqnarray}
Consequently, the initial decay rate is
\begin{eqnarray}
\Gamma=-\frac{\dot{\mathcal{N}}}{\mathcal{N}}\simeq \frac{1}{\mathcal{N}\Delta t}\simeq \frac{c}{\varphi_{end}\sqrt{3V_0\mathcal{N}}}\,,
\end{eqnarray}
so that 
\begin{eqnarray}
\bar{\varepsilon}\simeq \varepsilon_{\mathcal{N}}=\frac{\Gamma}{2H}\simeq\frac{c}{2\varphi_{end}V_0\mathcal{N}}\,. \label{barepsilonlinear}
\end{eqnarray}
Similarly, $\delta$ can be computed using $\dot{\Gamma}=\Gamma^2-\ddot{\mathcal{N}}/\mathcal{N}$ with $\ddot{\mathcal{N}}/\mathcal{N}\simeq 1/(2\mathcal{N}(\Delta t)^2)\simeq \Gamma^2/2$, so that
\begin{eqnarray}
\delta=\frac{\dot{\Gamma}H}{\Gamma \dot{H}}\simeq -\frac{\dot{\Gamma}}{2H^2}\frac{1}{\bar{\varepsilon}^2}\simeq -1\,.
\end{eqnarray} 
Further, the number of e-folds $N\approx 60$ becomes 
\begin{eqnarray}
N&=&\int_{ini}^{end} H\, dt 
\simeq -\int_{ini}^{end}\sum_{i=1}^{\mathcal{N}}\frac{V_i}{V_i^{\prime}}d\varphi_i\\
&\simeq& \frac{V_0\varphi_{end}}{c}\sum_{i=1}^{\mathcal{N}}\left(1-\frac{i-1}{\mathcal{N}}\right)\\
&\simeq&\frac{V_0\varphi_{end}\mathcal{N}}{2c} \,, \label{efoldslinear}
\end{eqnarray}
where we used the slow roll approximation in the first line, the initial conditions from (\ref{inicon}), $\tilde{\varepsilon}\ll 1 $ in the second one,  and the large $\mathcal{N}$ limit in the last one. We can rewrite this expression as $N\simeq \varphi_{end}\mathcal{N}/(2\tilde{\varepsilon})$. Thus, with about a thousand fields $\mathcal{N}\sim 10^3$ and flat potentials $\tilde{\varepsilon}\sim 10^{-2}$ we need the sharp drop to be around $\varphi_{end}\simeq 2\tilde{\varepsilon}N/\mathcal{N}\sim 3.5\times 10^{-2}\ll 1$ to yield around sixty e-folds of inflation. This is consistent with  our expansion, since potentials should be well behaved over such small stretches in field space. Further, our requirement of a drastic change in the potential (such as a sharp drop) once fields get to $\varphi_{end}$ appears natural.
Given(\ref{efoldslinear}), we can write
\begin{eqnarray}
\bar{\varepsilon}\simeq \frac{1}{4N}\,.
\end{eqnarray}
The resulting scalar spectral index (\ref{scalarspectralindex}), tensor spectral index (\ref{tensorspectralindex}) and tensor to scalar ratio (\ref{tensortoscalarratio}) are
\begin{eqnarray}
n_s-1&\simeq&-\frac{1}{N}\,,\label{nslinear}\\
n_T&\simeq&-\frac{1}{2N}\,,\\
r&\simeq&\frac{4}{N}\,. \label{rlinear}
\end{eqnarray}
Note that $r=-4(n_s-1)$, just like in a single-field model with $V=m^2\varphi^2/2$, but the values for $r$ and $n_s-1$ are smaller. A plot of the $r$-$n_s$ plane including the WMAP5 constraints \cite{Komatsu:2008hk} is in Fig.~\ref{pic:roverns}.

\subsubsection{Different Slopes \label{sec:linearv}}
\begin{figure}[tb]
\includegraphics[scale=0.88,angle=0]{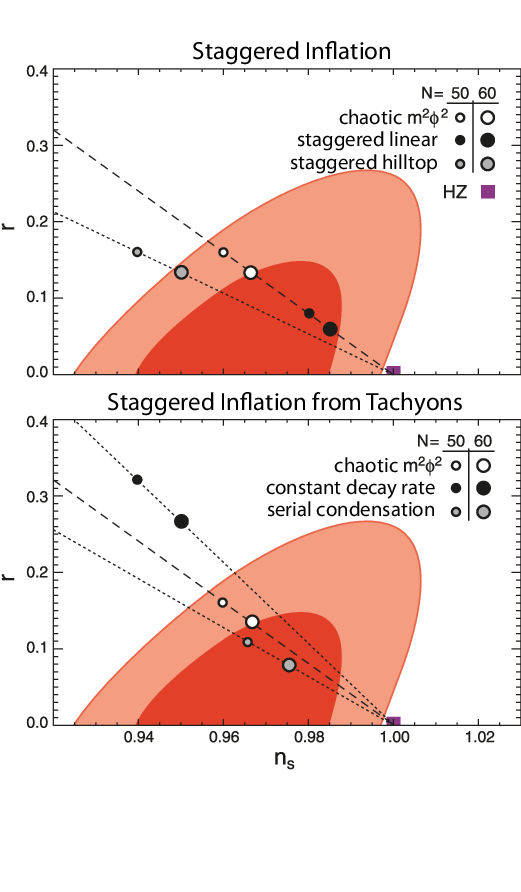}
   \caption{\label{pic:roverns} We compare the predictions of staggered inflation with the WMAP5 constraints (reproduced from \cite{Komatsu:2008hk} by permission of the AAS) and the canonical $m^2\varphi^2$ chaotic inflation model. Shaded areas correspond to $1\sigma$ and $2\sigma$ confidence levels. Top panel: models explored in this paper with linear potentials ($V_i=V_0-c \varphi_i$, section \ref{sec:linear}) or hilltop potentials ($V_i=V_0-m^2\varphi^2/2$, section \ref{sec:hilltop}); allowing for a spread of slopes or masses ($l\neq 0$), shifts predictions up along the dotted lines. The hilltop case is borderline ruled out at the $2\sigma$ level. Bottom panel: staggered inflation from multiple tachyons near a hilltop that condense due to fluctuations (not due to slow roll) as proposed in \cite{Majumdar:2003kd,Battefeld:2008py}, either with a constant decay rate ($\Gamma = constant$) or in a serial fashion ($\Gamma=(\tilde{t}-1)^{-1}$ so that  $\bar{\varepsilon}=1/(3N)$). The constant decay rate is already ruled out at the $2\sigma$ level, while serial condenstaion is still viable.}
\end{figure}

In realistic scenarios, it is unlikely that all potentials are identical, but we expect some spread of slopes. However, since fields with much larger slopes drop down quickly, we consider a narrow spread of the $c_i$'s only; specifically, we take
\begin{eqnarray}
c_i=c\left(1+l\frac{i}{\mathcal{N}}\right)\,,
\end{eqnarray}
with $l\ll 1$. If we use the same initial conditions as in (\ref{inicon}) with $\varphi_{ini}=0$ and use the above spread of $c_i$'s, we implicity assume that fields with a larger slope are already further down their potential. This seems reasonable to us, but the exact pairing of initial conditions and slopes could be chosen differently.  Based on our choice,  the number of e-folds becomes
\begin{eqnarray}
N\approx \frac{\varphi_{end}V_0\mathcal{N}}{2c}\left(1-\frac{l}{3}\right)\,, 
\end{eqnarray}
where we expanded for small $l$. Further, $\bar{\varepsilon}\simeq (1+2l/3)/(4N)$ and $\delta \simeq-1$, so that 
\begin{eqnarray}
n_s-1&\simeq&-\frac{1}{N}\left(1+\frac{2l}{3}\right)\,,\\
n_T&\simeq&-\frac{1}{2N}\left(1+\frac{2l}{3}\right)\,,\\
r&\simeq &\frac{4}{N}\left(1+\frac{2l}{3}\right)\,.
\end{eqnarray}
Thus, by increasing $l$, that is by spreading out the $c_i$'s, the scalar spectral index and tensor to scalar ratio shift up along the $r=-4(n_s-1)$ line in Fig.~\ref{pic:roverns}, closer to the canonical chaotic inflation case.

\subsection{Hilltop Inflation: $V_i=V_0-m^2\varphi_i^2/2$ \label{sec:hilltop}}
Consider a large number of fields in the vicinity of the maximum in their respective potentials.
If we expand around a maximum, we have to be careful to avoid initial values too close to the top: fluctuations dominate over the classical evolution of the fields if fields are too close to the maximum. Since $\delta \varphi_i^{QM}\sim H \sim 10^{-6}$ we take $\varphi_{ini}\equiv \varphi_{end}/\alpha$ with $\alpha=3.5 \times 10^{-4}$ (anticipating $\varphi_{end}\approx 3.5 \times 10^{-2}$ from below) in (\ref{inicon}), that is
\begin{eqnarray}
\varphi_{i}(0)=\frac{\varphi_{end}}{\alpha}\left(1+(\alpha-1)\frac{i-1}{\mathcal{N}}\right)\,.
\end{eqnarray}
As a result, $\varphi_i(0)-\varphi_{i-1}(0)\sim \varphi_{end}/\mathcal{N}\sim 3.5\times 10^{-5}>\varphi_{ini}$ where $\mathcal{N}\sim 10^{3}$, as it should. Given the potential and initial conditions, the slow roll equations of motion (\ref{slowroll}) integrate (for $\mathcal{N}\sim const$) to 
\begin{eqnarray}
\varphi_i(t)\simeq \varphi_i(0)e^{\frac{m^2}{\sqrt{3\mathcal{N}V_0}}t}\,.
\end{eqnarray}
Using this result, we can, once again, estimate the decay rate to
\begin{eqnarray}
\Gamma=-\frac{\dot{\mathcal{N}}}{\mathcal{N}}\simeq \frac{1}{\mathcal{N}\Delta t}\simeq \frac{m^2}
{\sqrt{3\mathcal{N}V_0}}\frac{\alpha}{\alpha-1}\,,
\end{eqnarray}
where 
\begin{eqnarray}
\Delta t \simeq \frac{\sqrt{3\mathcal{N}V_0}}{m^2}\ln\left(\frac{\alpha}{1+(\alpha-1)(1-1/\mathcal{N})}\right)\approx \frac{\alpha-1}{\alpha }\frac{1}{m^2}\sqrt{\frac{3V_0}{\mathcal{N}}}
\end{eqnarray}
is the time it takes for the $\mathcal{N}$'th field to encounter the drop; here, we only kept the leading order in 
$1/\mathcal{N}$ in the last step. Consequently,
\begin{eqnarray}
\bar{\varepsilon}\simeq \frac{m^2}{2V_0\mathcal{N}}\frac{\alpha}{\alpha-1}\,. \label{barvarepsilonhilltop}
\end{eqnarray}
Similarly, after some algebra, $\delta = \dot{\Gamma}H/(\Gamma \dot{H})$ becomes
\begin{eqnarray}
\delta\approx -3+\frac{2}{\alpha}\,,
\end{eqnarray}
to leading order in $1/\mathcal{N}$.
In what follows, we would like to express $\bar{\varepsilon}$ in terms of the number of e-folds $N$. The latter becomes 
\begin{eqnarray}
N\simeq -\int_{ini}^{end}\sum_{i=1}^{\mathcal{N}}\frac{V_i}{V_i^{\prime}}d\varphi_i= g(\alpha,\mathcal{N})\frac{V_0\mathcal{N}}{m^2}\,,
\end{eqnarray}
where $m^2/ V_0^2=2\tilde{\varepsilon}/\varphi_{end}^2$ and 
\begin{eqnarray}
g\equiv \ln(\alpha \mathcal{N})-\ln(\alpha-1)-\frac{1}{\mathcal{N}}\ln\left(\frac{\bar{\Gamma}(\mathcal{N}\alpha /(\alpha-1))}{\bar{\Gamma}(\mathcal{N}/(\alpha-1))}\right)\,.
\end{eqnarray}
Generically, $g$ is of order one for the models we are interested in and does not depend much on $\mathcal{N}$, for instance  we get $g\approx 1.0059$ for $\mathcal{N}\sim 10^3$ and $\alpha = 3.5\times  10^{4}$. As a consequence, we need again $\varphi_{end}\approx 2N\tilde{\varepsilon}/\mathcal{N}\approx 3.5\times 10^{-2}$ to achieve sixty e-folds of inflation. 
Thus, we have 
\begin{eqnarray}
\bar{\varepsilon}\simeq g\frac{\alpha}{\alpha-1}\frac{1}{2N}\approx \frac{1}{2N}
\end{eqnarray}
The resulting scalar spectral index (\ref{scalarspectralindex}), tensor spectral index (\ref{tensorspectralindex}) and tensor to scalar ratio (\ref{tensortoscalarratio}) are
\begin{eqnarray}
n_s-1&\simeq&\left(-6+\frac{2}{\alpha}\right)\frac{\alpha}{\alpha-1}\frac{g}{2N}\approx -\frac{3}{N}\,, \label{nsquadratic}\\
n_T&\simeq&-2 \frac{\alpha}{\alpha-1}\frac{g}{2N}\approx -\frac{1}{N}\,,\\
r&\simeq&16 \frac{\alpha}{\alpha-1}\frac{g}{2N} \approx \frac{8}{N}\,. \label{rquadratic}
\end{eqnarray}
Plotting the above in the $r$-$n_s$ plane including the WMAP5 constraints \cite{Komatsu:2008hk} in Fig.~\ref{pic:roverns} reveals that this model is already borderline ruled out at the $2\sigma$-level. 

\subsubsection{Different Masses \label{sec:quadraticv}}
Similar to Sec.~\ref{sec:linearv}, we would like to allow for slightly different potentials for each field, that is, slightly different masses. Choosing $m_i^2=m^2(1+l i/\mathcal{N})$ with $l\ll 1$ in conjunction with unchanged initial conditions, we implicitly assume that heavier fields are further down in the potential, which seems reasonable to us. 
Just as in the linear case, one can show that $\delta\approx -3$ remains unchanged to leading order in $1/\mathcal{N}$. Thus, predictions stay on the same $r(n_s)$ line in Fig.~\ref{pic:roverns}. Further, $\bar{\varepsilon}$ becomes
\begin{eqnarray}
\bar{\varepsilon}=\frac{m^2}{2V_0\mathcal{N}}(1+l)\frac{\alpha}{\alpha-1}\,,
\end{eqnarray}
and the number of e-folds reads
\begin{eqnarray}
N=\frac{V_0\mathcal{N}}{m^2}\tilde{g}(\alpha,\mathcal{N},l)\,,
\end{eqnarray}
where
\begin{eqnarray}
\tilde{g}=\frac{1}{\mathcal{N}}\sum_{i=1}^{\mathcal{N}}\frac{1}{1-li/\mathcal{N}}\ln\left(\frac{\varphi_{end}}{\varphi_i(0)}\right)\,.
\end{eqnarray}
The latter expression can be expanded to $\tilde{g}\approx 1.0059-0.2510 l\approx 1-l/4$ for $\mathcal{N}=1000$, $\alpha=3.5\times 10^4$ and $l \ll 1$. Thus
$\bar{\varepsilon}\approx (2N)^{-1}(1+3l/4)$
and as a consequence, predictions move further away from the desired region in Fig.~\ref{pic:roverns}, 
\begin{eqnarray}
n_s-1&\approx& -\frac{3}{N}\left(1+l\frac{3}{4}\right)\,, \\
n_T&\approx& -\frac{1}{N}\left(1+l\frac{3}{4}\right)\,,\\
r&\approx& \frac{8}{N}\left(1+l\frac{3}{4}\right)\,. 
\end{eqnarray}

\subsection{Discussion \label{sec:discussion}}
The predictions made in (\ref{nslinear})-(\ref{rlinear}) and (\ref{nsquadratic})-(\ref{rquadratic}) are generic for staggered inflation: the main assumption was that a large number of fields, $\mathcal{N}\sim \mathcal{O}(10^3)$, encounters a flat, sub-Planckian stretch in field space before the potential drops off again. Flat means that the standard slow roll parameters of assisted inflation and $\tilde{\varepsilon}=\Delta V_i/V_0$ are small \footnote{The latter parameter guarantees that $\varepsilon \ll \bar{\varepsilon}$, so that effects due to staggered inflation dominate; otherwise, one would have a standard assisted inflation setup.}. We expanded the potentials, keeping only the linear or quadratic terms, depending on whether the fields are close to maxima or generic flat stretches. Since only fields with similar shallow potentials contribute to inflation for an extended period, we focused on fields whose potentials are similar. After distributing the fields evenly, which appears to be a natural starting point, we observed that the slow roll dynamics causes fields to decay during inflation, one after the other; this determines $\bar{\varepsilon}$ as well as $\delta$, which are not free parameters. The scalar and tensor spectral indices, as well as the tensor to scalar ratio follow directly and can be expressed in terms of the number of e-folds $N\approx 60$. 

The hilltop case is already borderline ruled out at the $2\sigma$ level according to the WMAP5 constraints \cite{Komatsu:2008hk}, see Fig.~\ref{pic:roverns} top panel. If we allow for a slight spread of masses as in Sec.~\ref{sec:quadraticv}, while retaining the same initial conditions, the predicted values of $r$ and $n_s$ move up along the $r(n_s)$-line, further away from the desired region. Thus, the only way to reconcile a hilltop model with observations in staggered inflation relies in fine tuned initial conditions, quite different from the even distribution we used. The more generic linear case (see Sec. \ref{sec:landscape}) is still viable. For comparison, we plot a single-field model with $m^2\varphi^2$ potential. Note that both, staggered linear inflation and chaotic $m^2\varphi^2$ inflation, have $r=-4(n_s-1)$, but the magnitude of $r$ and $|n_s-1|$ is smaller in staggered inflation. However, if we allow for different slopes as in Sec.~\ref{sec:linearv}, the predictions move up along the $r(n_s)$-line, closer to the canonical chaotic inflation case. Thus, the two models become degenerate in this plot, but could be discriminated by an observation of $n_T$.

An advantage of staggered inflation with the chosen parameters over such a large-field model is the possible avoidance of the $\eta$-problem \cite{Liddle:1998jc,Kanti:1999vt,Kanti:1999ie}, since no field has to traverse a distance longer than $\varphi_{end}\sim 3.5\times 10^{-2}\ll 1$. If we either increase the number of fields, or we decrease $\tilde{\varepsilon}$, $\varphi_{end}$ becomes smaller, without affecting the above observables.

\subsubsection{Concrete Model: Inflation from Tachyons}
In the bottom panel of Fig.~\ref{pic:roverns} we plot $n_s$ and $r$ for staggered inflation from multiple tachyons, a  model proposed in \cite{Battefeld:2008py,Majumdar:2003kd}. From a higher dimensional point of view, the tachyons are related to brane anti-brane pairs ($Dp-\bar{D}p$). By focussing on the Abelian part of the $U(\mathcal{N})\times U(\mathcal{N})$ gauge symmetry, Davis and Majumdar arrived at $\mathcal{N}$ tachyons with potential (see \cite{Ohmori:2001am} for a derivation as well as a general review on tachyons)
\begin{eqnarray}
W=\mathcal{N}\tau_p -c_1\sum_{i=1}^{\mathcal{N}}|\varphi_i|^2+c_2\sum_{i=1}^{\mathcal{N}}|\varphi_i|^4+\mathcal{O}\left(|\varphi_i|^6\right)\,. \label{potentialacd}
\end{eqnarray}
This potential is only valid in close proximity to $(\varphi_1,...,\varphi_{\mathcal{N}})=0$. Here $\tau_p$ is a model dependent brane tension of order one while $c_1\approx 0.87$ and $c_2\approx 0.21$ \cite{Ohmori:2001am}. The hundreds of tachyons start out close to the top of their potentials and they condense, that is decay, in a staggered fashion due to fluctuations, not because of slow roll. Thus, one can give all fields identical initial values without loss of generality. 

We arrive at the same expressions for $n_s$, $r$ and $n_T$ as in (\ref{scalarspectralindex}), (\ref{tensortoscalarratio}) and (\ref{tensorspectralindex}), if the slow roll contributions are negligible; disregarding these contributions is well motivated within this scenario, because the employed potential ceases to be valid before $\varepsilon$ becomes of order $\bar{\varepsilon}$ \cite{Battefeld:2008py}  \footnote{The tensor to scalar ratio reads $r= 16(\varepsilon \gamma^2+\bar{\varepsilon})$ if slow roll contributions were included; here, $\gamma$ is a parameter of order one that can be determined from the potential as in \cite{Battefeld:2008py}.}. Note that we treat the decay rate as a free function, because we lack a good understanding of tachyon condensation \footnote{As a consequence, (p)reheating is presently not understood in this proposal and might indeed be problematic due to couplings to hidden sectors.}; two simple choices for $\Gamma$ where proposed \cite{Majumdar:2003kd,Battefeld:2008py}: first, a {\em constant decay rate} $\Gamma=const$ so that $\delta=0$ and $\bar{\varepsilon}=1/N$; second, $\Gamma=(\tilde{t}-t)^{-1}$ where $\tilde{t}=const$ leading to {\em serial condensation} so that $\delta =-2$ and $\bar{\varepsilon}=1/(3N)$. The constant decay rate is already ruled out at the $2\sigma$ level, while serial condensation of tachyons provides a model in good agreement with observations Fig.~\ref{pic:roverns}. For further constraints on $\mathcal{N}$ and $\tau_p$, based on the requirement of achieving at least sixty efolds of inflation, we refer the interested reader to \cite{Battefeld:2008py}.  

\subsubsection{Summary}
To summarize, current observational bounds already rule out several models of staggered inflation driven by fields close to maxima in their potentials, albeit a model with tachyons that condense serially is still in good agreement with observations. Also, the most generic case is still viable: many fields roll along flat stretches that can be approximated by linear potentials up until $\Delta \varphi_i \sim 10^{-2}$ where a sharp drop follows. In the next section, we elaborate more on this model and speculate how it could arise in a realistic model on the landscape.

\section{Staggered Inflation on the Landscape? \label{sec:landscape}}
The models studied so far appear to have ample freedom regarding the slope of potentials as well as initial conditions. If one has the freedom to choose either of them at will, one clearly looses any predictive power. In order to make the precise predictions of the last sections, we employ linear or hilltop potentials with a narrow distribution of slopes/masses as well as an even distribution of initial field values over the allowed interval. Here, we would like to re-examine how natural, or fine tuned, our preferences are.

First of all, we need around a thousand fields for staggered inflation to work. Such a large number seems intimidating at first, but it is actually in the right order of magnitude for string theory, which contains many degrees of freedom (shape and size moduli, fluxes, branes, etc.). Instead of stabilizing all but one degree of freedom, as is usually done in attempts of implementing inflation into string theory \cite{McAllister:2007bg,Cline:2006hu}, we let them evolve during inflation. This is similar in spirit to $\mathcal{N}$-flation \cite{Dimopoulos:2005ac} ($\sim 1000$ axions), inflation from multiple $M5$ branes \cite{Becker:2005sg} ($\sim 100$ fields) or chain inflation \cite{Freese:2004vs}. Indeed, the mere fact that many degrees of freedom are involved in inflation appears less fine tuned to us.   

How natural are the potentials we consider? We primarily assumed that there are some flat stretches over short intervals $\Delta \varphi_i \ll 1$ followed by sharp drops with large slopes. Such potentials are less fine tuned than the ones employed in large-field models with a single inflaton field: in these setups potentials need to be flat over $\Delta \varphi \sim 10$, which is hard to achieve, since quantum corrections usually lead to large contributions once $\Delta \varphi \sim 1$ (the $\eta$ problem). Stabilizing all but one modulus and the necessity of keeping corrections under control are the reasons that single-field models like the KKLMMT proposal \cite{Kachru:2003sx} appear rather baroque \footnote{These models have the advantage that all moduli are guaranteed to be securely fixed at late times, as they should be. See \cite{Baumann:2008kq,Baumann:2007ah} for estimates of corrections to the inflaton potential.}.

In contrast, we rely on large corrections to potentials that cause a given field to evolve fast once $\Delta \varphi_i$ becomes of order $10^{-2}$. All we assume is that the subsequent evolution is rapid, leading to stabilization or decay of the field while releasing energy in the form of new degrees of freedom, e.g. radiation. Thus, we incorporate the notion of moduli trapping on the landscape \cite{Kofman:2004yc,Watson:2004aq}, but instead of assuming that the stabilization epoch ends before inflation commences, we propose that fields are still stabilizing during the last sixty e-folds. The completion of moduli stabilization goes hand in hand with the end of inflation and reheating, tying together two previously distinct phenomena. In this sense, staggered multi-field inflation is more economic than single-field models which require an additional (unobservable) earlier phase of moduli stabilization.

Let's turn our attention to initial conditions; we employ an even distribution over regions where the potential is flat. How generic is this distribution? Addressing this question is challenging, due to our inability to assign a meaningful measure on the space of initial conditions (or the landscape in general for that matter; see i.e. \cite{Linde:2007nm,Gibbons:2006pa,Hartle:2007gi} for different proposals, \cite{Linde:2007fr} for a short review and \cite{Denef:2008wq} for more comprehensive lecture notes). Nevertheless, we can give some plausible, heuristic arguments: consider a pre-inflationary phase where the potential energy is negligible compared to the kinetic energy of fields \footnote{If potentials are unbounded from above, they would introduce hard walls from which the fields get reflected, restricting the accessible regions on the landscape. There are models employing such features, such as the new-ekpyrotic scenario, see \cite{Lehners:2008vx} for a review.}. Such a regime arises naturally after a Hagedorn phase  \footnote{A phase where an increase of energy does not cause a corresponding  increase in the temperature, but the excitations of long strings. This feature of a limiting Hagedorn temperature \cite{Hagedorn:1965st} is a generic feature of string theory, which encompasses T-duality.}, which has recently gained a lot of interest in String Gas Cosmology and bouncing models of the Universe (see e.g. \cite{Novello:2008ra,Battefeld:2005av,Brandenberger:2008nx} for reviews). Since the fields move freely, they randomly scan all regions of the landscape accessible to them. While the universe is expanding (non inflationary), the kinetic energy redshifts until the potential energy becomes comparable to the kinetic one. At this instance, fields are at random positions on the landscape, a few already close to a minimum (or maximum), but the vast majority of fields find themselves on intermediate stretches that have non-zero slope. If the slope on such a stretch is large, the field relaxes quickly to a minimum (it gets stabilized/decays), but if the slope is flat enough, the evolution becomes over-damped and the field enters a slow roll phase. After a few Hubble times, the fields with the smallest slopes dominate the energy density of the universe and start driving assisted inflation \cite{Liddle:1998jc}. Of course, only those with comparable (large) energy densities need to be considered here. Thus, we arrive dynamically at a setup similar to the one considered in this article, starting from a generic initial state. 

It seems plausible that intermediate stretches occupy more ground on the landscape than the very flat regions near extrema; hence, we expect to find the majority of fields in regions where linear potentials serve as good approximations. Thus, this line of reasoning favors the linear case over hilltop potentials \footnote{Classically, fields extremely close to a maximum would be the ones which survive the longest, and hence one might be tempted to conclude that those fields are the ones most likely to drive the last sixty e-folds of inflation. However, quantum mechanical fluctuations displace these fields so that they start rolling classically within a single Hubble time. They will then encounter cliffs on comparable time scales as fields on linear stretches. Thus, it is unlikely that a large number of fields simply hangs around near a maximum until late times.}; this is good news, since we saw that hilltop potentials are already ruled out at the $2\sigma$-level. However, since we do not have a canonical measure on the landscape, these statements are mathematically ill defined. Therefore, we examined both cases in this paper. Nevertheless, if the fields have a periodic field domain, which is the case if the inflatons were for instance identified with axions as in $\mathcal{N}$-flation, one could make the above statements more concrete, since measures can be defined properly (we leave this to a future study). We anticipate that one is naturally led to an even distribution over the entire (finite) domain just before the fields start to see their potential. In this light, our choice of an even initial field-distribution is well motivated.

\section{Outlook\label{sec:opensissues}}
Here, we summarize the main open questions regarding staggered inflation.

\paragraph{Validity of the Analytic Framework:} our predictions rely on taking a continuum limit that leads to a smooth decay rate $\Gamma$. This approximation becomes exact in the limit of $\mathcal{N}\rightarrow \infty$ while keeping $W$ and $\Gamma$ fixed, but it remains to be seen how low $\mathcal{N}$ and $\Gamma$ can be, before effects caused by the stabilization/decay of individual fields become observable. The physical reason for additional effects is the violation of slow roll by the decaying fields, causing i.e. a ringing in the scalar power-spectrum, similar to the ringing after a sharp step in the potential of a single inflaton field \cite{Chen:2006xjb,Chen:2008wn}; in \cite{Ashoorioon:2008qr} a simplified two field model was investigated numerically and indeed a ringing was found, but it turned out to be damped more than in the single-field case. If many fields decay in a given Hubble time, those features will be washed out even further, but exactly how many fields are needed is unknown. A numerical treatment should be performed, but a better understanding of the fields' decay is needed beforehand. 

\paragraph{The Nature of the Fields' Decay and Reheating:} so far, an efficient and fast energy transfer from the decaying/stabilizing inflatons to radiation or other types of matter with a constant equation of state is assumed. The exact nature of this process is model dependent and differs considerably from case to case. It would be instructive to have a concrete toy model, enabling the numerical check of the analytic framework; the extensive literature on different reheating mechanisms after inflation (see e.g. \cite{Bassett:2005xm} for a review) should be a valuable resource in this endeavor. Further, given a concrete model, the computation of additional contributions to obervables, i.e. non-Gaussianities and gravitational waves, becomes possible.

But is reheating of standard model particles for the model at hand feasable in the first place? Due to the presence of a large number of fields, many of which  are conceivably coupled to hidden sectors, one might fear that reheaing hidden sectors is a severe problem, just like in $\mathcal{N}$-flation \cite{Green:2007gs}. Fortunately, there is a promising way of incorporating reheating into the framework of staggered inflation: since the fields decay one after the other, only the field(s) that decay last need to decay into (extensions of the) standard model degrees of freedom. Since we considered models on the landscape, we should include all possible degrees of freedom, among which there will be i.e. those that were used in \cite{Allahverdi:2006iq,Allahverdi:2006cx,Allahverdi:2006we,Allahverdi:2008bt} to drive inflation with the field content of the MSSM \footnote{This line of reasoning is one of the main motivations for implementing Inflation within the MSSM; note that in order to realize the low-scale inflation of i.e. \cite{Allahverdi:2006iq,Allahverdi:2006cx,Allahverdi:2006we,Allahverdi:2008bt} one generically needs a prior high-scale inflationary era of the type we consider. Hence, low- and high-scale inflation complement each other.} (see also \cite{Jokinen:2004bp} for a multi-field model). If the last field to decay is of this type, standad model degrees of freedom would be predominantly reheated \cite{Allahverdi:2006we}, for instance via instant preheating \cite{Felder:1998vq,Allahverdi:2006we}. The decay products of the previous fields (which are most likely  in hidden sectors) would have been diluted to some degree at the end of inflation and might very well account for dark matter \footnote{Depending on
the concrete implementation of reheating standard model degrees of freedom, observables such as the scalar spectral index might differ from the ones predicted here: for instance, if a staggered inflation is followed by an extended low-scale, single-field inflationary phase driven \cite{Allahverdi:2006iq,Allahverdi:2006cx,Allahverdi:2006we,Allahverdi:2008bt} that lasts sufficiently long, the power spectrum could be entirely determined by the latter. However, even in this case an understanding of the preceding staggered inflationary phase is useful, since it determines the initial state of the later low-scale inflationary stage. A study of reheating after multi-field inflation is in progress \cite{inprep2}.}. Another option would be to reheat via MSSM flat directions (see i.e. \cite{Enqvist:2003qc} and follow ups).

\paragraph{Isocurvature Perturbations and Non-Gaussianities:} within the analytic framework, isocurvature perturbations are suppressed \cite{Battefeld:2008py}, leading to the expectation that non-Gaussianities are suppressed too \cite{Battefeld:2008py}, just as in models of multi-field slow roll inflation \cite{Battefeld:2006sz,Battefeld:2007en}. Deviations from this expectation are feasible during the short instances when fields decay, leading to the production of non-Gaussianities similar to the ones generated in single-field models with a sharp step in the potential \cite{Chen:2006xjb,Chen:2008wn}, or during (p)re-heating \cite{Barnaby:2006cq,Barnaby:2006km,Chambers:2008gu,Enqvist:2004ey,Enqvist:2005qu,Jokinen:2005by}. Since fields are decaying/stabilizing throughout inflation, sizeable non-Gaussianities might result, which could very well dominate over primordial ones. 

\paragraph{Additional Gravitational Waves:} it is known that gravitational waves can be produced on sub-Hubble scales during (p)reheating (for a selection of numerical studies see \cite{Khlebnikov:1997di,GarciaBellido:1998gm,Easther:2006gt,GarciaBellido:2007dg,GarciaBellido:2007af,Dufaux:2007pt,Easther:2007vj,Easther:2006vd}). Because fields decay throughout inflation, we expect an additional gravitational wave spectrum resulting from the superposition of spectra similar to the ones found in \cite{Easther:2007vj}; since the process operates within the horizon, the peaks of the spectra are always generated at the same fraction of the Hubble radius whenever a field decays. Once produced, inflation stretches them to super Hubble scales, just as it does with gravity waves seeded by quantum fluctuations. Further, since each decay resembles the previous ones, the amplitude at which gravity waves are produced should not change either. As a consequence, the resulting superposition of power-spectra should lead to a spectrum with a spectral index quite different from $n_T\simeq -2\bar{\varepsilon}$. Whether the primordial powerspectrum or the one from decaying fields dominates, dependents on the detailed physics of the fields' decay mechanism. Once a concrete model within string theory is constructed and the dacays are better understood, the additional contribution to $\mathcal{P}_T$ is computable and could serve as a smoking gun for staggered multi-field inflation.

\paragraph{Feasibility of Staggered Inflation on the Landscape:} the arguments within this paper on the occurrence of staggered inflation on the landscape were heuristic, due to the fact that a canonical measure on the landscape is still missing. Only after such a measure is known, one can properly estimate whether or not staggered inflation is generic, and whether or not initial conditions are fine tuned. Nevertheless, the heuristic arguments put forward in this paper are promising.

\paragraph{Generalizations:} so far, staggered inflation has been examined for separable potentials, canonical kinetic terms and a flat field-space metric only. String theory on the other hand permits more general scenarios; i.e. it would be interesting to study the effects of decaying fields during DBI inflation \cite{Silverstein:2003hf,Alishahiha:2004eh,Chen:2004gc,Chen:2005ad,Easson:2007dh,Langlois:2008wt,Langlois:2008qf} or k-inflation \cite{ArmendarizPicon:1999rj,Garriga:1999vw}.

\section{Conclusions \label{sec:conclusion}}
Staggered multi-field inflation incorporates the notion of decaying/stabilizing fields during inflation, a feature that has been largely ignored in the literature. We extended the formalism to incorporate more general initial conditions and potentials, enabling the discussion of generic multi-field models as expected in the landscape picture. The potentials we considered are flat over short intervals, before a sharp drop is encountered. Such potentials are less fine tuned than the one in single-field models, which require flatness over long distances in field space. After spreading out the fields evenly over the allowed interval and using either linear or hilltop potentials with some spread of slopes or masses, we computed observable parameters such as the scalar spectral index, the tensor to scalar ratio or the tensor spectral index. These parameters are dominated by the contributions due to staggered inflation $\bar{\varepsilon} \simeq \Gamma / (2H) \propto \rho_{radiation}/\rho_{total} > \varepsilon_{slowroll}$; $\bar{\varepsilon}$ is not a free parameter but follows from the slow roll evolution.

We found that hilltop potentials are already ruled out at the $2\sigma$ level by the WMAP5 data, whereas linear potentials, which are better motivated, are in excellent agreement with observations. We also reconsidered staggered inflation driven by multiple tachyons \cite{Majumdar:2003kd,Battefeld:2008py}, where $\Gamma(t)$ is an unknown function. We find that constant decay rates are already ruled out at the $2\sigma$ level, but a serial condensation of tachyons is still viable.  

One drawback of multi-field inflation is the apparent flexibility to choose initial values and potentials at will, increasing the tunability compared to single-field scenarios. As a result, multi-field models are in danger of loosing their predictability. To minimize fine tuning, we investigated which potentials and initial conditions appear to be generic on the landscape and gave arguments in favor of our choice. However, the absence of a canonical measure on the space of initial conditions and the landscape renders these arguments heuristic and open to debate. This uncertainty can be alleviated in certain models, for instance if the inflatons were axions with a finite domain. 

We concluded with a summary of open questions, ranging from a numerical validation of the analytic framework, to potentially new sources of gravitational waves and non-Gaussianities; the latter ones may very well serve as a smoking gun for staggered inflation, but require a better understanding of the fields' decay/stabilization mechanism before they can be computed. In this light, a more concrete implementation within string theory of this new type of inflation on the landscape is of great interest.  

%%%%%%%%%%%%%%%%%%%%%%%%%%%%%%%%%%%%
\begin{acknowledgments}
We thank A.~Ashoorioon, K.~Enqvist, D.~Langlois, E.~Lim, A.~Mazumdar and P.~Steinhardt for discussions. T.~B. is greateful for the hospitality at the Helsinki Institute of Physics and the APC (CNRS-Universit´e Paris 7). T.B. is supported by the Council on Science and Technology at Princeton University. D.~B is supported by the EU EP6 Marie Curie Research and Training Network 'UniverseNet' (MRTN-CT-2006-035863).

\end{acknowledgments}
%%%%%%%%%%%%%%%%%%%%%%%%%%%%%%%%%%%%

\appendix


\begin{thebibliography}{}
  
  %\cite{McAllister:2007bg}
\bibitem{McAllister:2007bg}
  L.~McAllister and E.~Silverstein,
  ``String Cosmology: A Review,''
  Gen.\ Rel.\ Grav.\  {\bf 40}, 565 (2008)
  [arXiv:0710.2951 [hep-th]].
  %%CITATION = GRGVA,40,565;%%
  
%\cite{Cline:2006hu}
\bibitem{Cline:2006hu}
  J.~M.~Cline,
  ``String cosmology,''
  arXiv:hep-th/0612129.
  %%CITATION = HEP-TH/0612129;%%
 
 %\cite{Burgess:2007pz}
\bibitem{Burgess:2007pz}
  C.~P.~Burgess,
  ``Lectures on Cosmic Inflation and its Potential Stringy Realizations,''
  PoS {\bf P2GC}, 008 (2006)
  [Class.\ Quant.\ Grav.\  {\bf 24}, S795 (2007)]
  [arXiv:0708.2865 [hep-th]].
  %%CITATION = CQGRD,24,S795;%%
  
  
%\cite{Kallosh:2007ig}
\bibitem{Kallosh:2007ig}
  R.~Kallosh,
  ``On Inflation in String Theory,''
  Lect.\ Notes Phys.\  {\bf 738}, 119 (2008)
  [arXiv:hep-th/0702059].
  %%CITATION = LNPHA,738,119;%%

%\cite{Kachru:2003sx}
\bibitem{Kachru:2003sx}
  S.~Kachru, R.~Kallosh, A.~Linde, J.~M.~Maldacena, L.~P.~McAllister and S.~P.~Trivedi,
  ``Towards inflation in string theory,''
  JCAP {\bf 0310}, 013 (2003)
  [arXiv:hep-th/0308055].
  %%CITATION = JCAPA,0310,013;%%
  


%\cite{Baumann:2007ah}
\bibitem{Baumann:2007ah}
  D.~Baumann, A.~Dymarsky, I.~R.~Klebanov and L.~McAllister,
  ``Towards an Explicit Model of D-brane Inflation,''
  JCAP {\bf 0801}, 024 (2008)
  [arXiv:0706.0360 [hep-th]].
  %%CITATION = JCAPA,0801,024;%%


  
  %\cite{Baumann:2008kq}
\bibitem{Baumann:2008kq}
  D.~Baumann, A.~Dymarsky, S.~Kachru, I.~R.~Klebanov and L.~McAllister,
  ``Holographic Systematics of D-brane Inflation,''
  JCAP {\bf 0801}, 24 (2008)
  [arXiv:0808.2811 [hep-th]].
  %%CITATION = JCAPA,0801,24;%%
  
%\cite{Hoi:2008gc}
\bibitem{Hoi:2008gc}
  L.~Hoi and J.~M.~Cline,
  ``How Delicate is Brane-Antibrane Inflation?,''
  arXiv:0810.1303 [hep-th].
  %%CITATION = ARXIV:0810.1303;%%  
  
%\cite{Liddle:1998jc}
\bibitem{Liddle:1998jc}
  A.~R.~Liddle, A.~Mazumdar and F.~E.~Schunck,
  ``Assisted inflation,''
  Phys.\ Rev.\  D {\bf 58}, 061301 (1998)
  [arXiv:astro-ph/9804177].
  %%CITATION = PHRVA,D58,061301;%%
  
   %\cite{Malik:1998gy}
\bibitem{Malik:1998gy}
  K.~A.~Malik and D.~Wands,
  ``Dynamics of assisted inflation,''
  Phys.\ Rev.\  D {\bf 59}, 123501 (1999)
  [arXiv:astro-ph/9812204].
  %%CITATION = PHRVA,D59,123501;%%
  
  %\cite{Kanti:1999vt}
\bibitem{Kanti:1999vt}
  P.~Kanti and K.~A.~Olive,
  ``On the realization of assisted inflation,''
  Phys.\ Rev.\ D {\bf 60}, 043502 (1999)
  [arXiv:hep-ph/9903524].
  %%CITATION = HEP-PH 9903524;%%
  
  %\cite{Kanti:1999ie}
\bibitem{Kanti:1999ie}
  P.~Kanti and K.~A.~Olive,
  ``Assisted chaotic inflation in higher dimensional theories,''
  Phys.\ Lett.\  B {\bf 464}, 192 (1999)
  [arXiv:hep-ph/9906331].
  %%CITATION = PHLTA,B464,192;%%
  
  
%\cite{Calcagni:2007sb}
\bibitem{Calcagni:2007sb}
  G.~Calcagni and A.~R.~Liddle,
  ``Stability of multi-field cosmological solutions,''
  Phys.\ Rev.\  D {\bf 77}, 023522 (2008)
  [arXiv:0711.3360 [astro-ph]].
  %%CITATION = PHRVA,D77,023522;%%

\bibitem{Dimopoulos:2005ac}
  S.~Dimopoulos, S.~Kachru, J.~McGreevy and J.~G.~Wacker,
  ``N-flation,''
  JCAP {\bf 0808}, 003 (2008)
  [arXiv:hep-th/0507205].
  %%CITATION = JCAPA,0808,003;%%
   

%\cite{Easther:2005zr}
\bibitem{Easther:2005zr}
  R.~Easther and L.~McAllister,
  ``Random matrices and the spectrum of N-flation,''
  JCAP {\bf 0605}, 018 (2006)
  [arXiv:hep-th/0512102].
  %%CITATION = HEP-TH 0512102;%%
  
%\cite{Kachru:2003aw}
\bibitem{Kachru:2003aw}
  S.~Kachru, R.~Kallosh, A.~Linde and S.~P.~Trivedi,
  ``De Sitter vacua in string theory,''
  Phys.\ Rev.\  D {\bf 68}, 046005 (2003)
  [arXiv:hep-th/0301240].
  %%CITATION = PHRVA,D68,046005;%%
  
     %\cite{Misra:2007cq}
\bibitem{Misra:2007cq}
  A.~Misra and P.~Shukla,
  ``Large Volume Axionic Swiss-Cheese Inflation,''
  Nucl.\ Phys.\  B {\bf 800}, 384 (2008)
  [arXiv:0712.1260 [hep-th]].
  %%CITATION = NUPHA,B800,384;%%

%\cite{Misra:2008tx}
\bibitem{Misra:2008tx}
  A.~Misra and P.~Shukla,
  ``'Finite' Non-Gaussianities and Tensor-Scalar Ratio in Large Volume
  Swiss-Cheese Compactifications,''
  arXiv:0807.0996 [hep-th].
  %%CITATION = ARXIV:0807.0996;%% 
  
   %\cite{Becker:2005sg}
\bibitem{Becker:2005sg}
  K.~Becker, M.~Becker and A.~Krause,
  ``M-theory inflation from multi M5-brane dynamics,''
  Nucl.\ Phys.\  B {\bf 715}, 349 (2005)
  [arXiv:hep-th/0501130].
  %%CITATION = NUPHA,B715,349;%%
  
  %\cite{Piao:2002vf}
\bibitem{Piao:2002vf}
  Y.~S.~Piao, R.~G.~Cai, X.~m.~Zhang and Y.~Z.~Zhang,
  ``Assisted tachyonic inflation,''
  Phys.\ Rev.\  D {\bf 66}, 121301 (2002)
  [arXiv:hep-ph/0207143].
  %%CITATION = PHRVA,D66,121301;%%


 %\cite{Majumdar:2003kd}
\bibitem{Majumdar:2003kd}
  M.~Majumdar and A.~C.~Davis,
  ``Inflation from tachyon condensation, large N effects,''
  Phys.\ Rev.\  D {\bf 69}, 103504 (2004)
  [arXiv:hep-th/0304226].
  %%CITATION = PHRVA,D69,103504;%%
  
   %\cite{Wands:2007bd}
\bibitem{Wands:2007bd}
  D.~Wands,
  ``Multiple field inflation,''
  Lect.\ Notes Phys.\  {\bf 738}, 275 (2008)
  [arXiv:astro-ph/0702187].
  %%CITATION = LNPHA,738,275;%%
  
    %\cite{Green:2007gs}
\bibitem{Green:2007gs}
  D.~R.~Green,
  ``Reheating Closed String Inflation,''
  Phys.\ Rev.\  D {\bf 76}, 103504 (2007)
  [arXiv:0707.3832 [hep-th]].
  %%CITATION = PHRVA,D76,103504;%%
    
  %\cite{Krause:2007jr}
\bibitem{Krause:2007jr}
  A.~Krause,
  ``Large Gravitational Waves and Lyth Bound in Multi Brane Inflation,''
  JCAP {\bf 0807}, 001 (2008)
  [arXiv:0708.4414 [hep-th]].
  %%CITATION = JCAPA,0807,001;%%


    %\cite{Ashoorioon:2006wc}
\bibitem{Ashoorioon:2006wc}
  A.~Ashoorioon and A.~Krause,
  ``Power spectrum and signatures for cascade inflation,''
  arXiv:hep-th/0607001.
  %%CITATION = HEP-TH/0607001;%%

  %\cite{Ashoorioon:2008qr}
\bibitem{Ashoorioon:2008qr}
  A.~Ashoorioon, A.~Krause and K.~Turzynski,
  ``Energy Transfer in Multi Field Inflation and Cosmological Perturbations,''
  arXiv:0810.4660 [hep-th].
  %%CITATION = ARXIV:0810.4660;%%
    
%\cite{Battefeld:2008py}
\bibitem{Battefeld:2008py}
  D.~Battefeld, T.~Battefeld and A.~C.~Davis,
  ``Staggered Multi-Field Inflation,''
  JCAP {\bf 0810}, 032 (2008)
  [arXiv:0806.1953 [hep-th]].
  %%CITATION = JCAPA,0810,032;%%
  
  
  %\cite{Battefeld:2008ur}
\bibitem{Battefeld:2008ur}
  T.~Battefeld,
  ``Exposition to Staggered Multi-Field Inflation,''
  arXiv:0809.3242 [astro-ph].
  %%CITATION = ARXIV:0809.3242;%%
  
  
  
  
%\cite{Battefeld:2008bu}
\bibitem{Battefeld:2008bu}
  D.~Battefeld and S.~Kawai,
  ``Preheating after N-flation,''
  Phys.\ Rev.\  D {\bf 77}, 123507 (2008)
  [arXiv:0803.0321 [astro-ph]].
  %%CITATION = PHRVA,D77,123507;%%
  
  %\cite{Battefeld:2008rd}
\bibitem{Battefeld:2008rd}
  D.~Battefeld,
  ``Preheating after Multi-field Inflation,''
  arXiv:0809.3455 [astro-ph].
  %%CITATION = ARXIV:0809.3455;%%

  
   
  %\cite{Dufaux:2006ee}
\bibitem{Dufaux:2006ee}
  J.~F.~Dufaux, G.~N.~Felder, L.~Kofman, M.~Peloso and D.~Podolsky,
  ``Preheating with trilinear interactions: Tachyonic resonance,''
  JCAP {\bf 0607}, 006 (2006)
  [arXiv:hep-ph/0602144].
  %%CITATION = JCAPA,0607,006;%%

 
\bibitem{inprep2} 
 D.~Battefeld, T.~Battefeld and J.~T.~Giblin, in preparation.
  
    
  %\cite{Branden}  
\bibitem{Braden} 
  J.~Braden, work presented at COSMO08, unpublished.

%\cite{Berera:2008ar}
\bibitem{Berera:2008ar}
  A.~Berera, I.~G.~Moss and R.~O.~Ramos,
  ``Warm Inflation and its Microphysical Basis,''
  arXiv:0808.1855 [hep-ph].
  %%CITATION = ARXIV:0808.1855;%%
  
    %\cite{Berera:1995ie}
\bibitem{Berera:1995ie}
  A.~Berera,
  ``Warm Inflation,''
  Phys.\ Rev.\ Lett.\  {\bf 75}, 3218 (1995)
  [arXiv:astro-ph/9509049].
  %%CITATION = PRLTA,75,3218;%%

    %\cite{Hall:2007qw}
\bibitem{Hall:2007qw}
  L.~M.~H.~Hall and H.~V.~Peiris,
  ``Cosmological Constraints on Dissipative Models of Inflation,''
  JCAP {\bf 0801}, 027 (2008)
  [arXiv:0709.2912 [astro-ph]].
  %%CITATION = JCAPA,0801,027;%%
  
  %\cite{Yokoyama:1998ju}
\bibitem{Yokoyama:1998ju}
  J.~Yokoyama and A.~D.~Linde,
  ``Is warm inflation possible?,''
  Phys.\ Rev.\  D {\bf 60}, 083509 (1999)
  [arXiv:hep-ph/9809409].
  %%CITATION = PHRVA,D60,083509;%%  
  
  %\cite{Susskind:2003kw}
\bibitem{Susskind:2003kw}
  L.~Susskind,
  ``The anthropic landscape of string theory,''
  arXiv:hep-th/0302219.
  %%CITATION = HEP-TH/0302219;%%

  %\cite{Freese:2004vs}
\bibitem{Freese:2004vs}
  K.~Freese and D.~Spolyar,
  ``Chain inflation: 'Bubble bubble toil and trouble',''
  JCAP {\bf 0507}, 007 (2005)
  [arXiv:hep-ph/0412145].
  %%CITATION = JCAPA,0507,007;%%

%\cite{Feldstein:2006hm}
\bibitem{Feldstein:2006hm}
  B.~Feldstein and B.~Tweedie,
  ``Density perturbations in chain inflation,''
  JCAP {\bf 0704}, 020 (2007)
  [arXiv:hep-ph/0611286].
  %%CITATION = JCAPA,0704,020;%%


%\cite{Freese:2006fk}
\bibitem{Freese:2006fk}
  K.~Freese, J.~T.~Liu and D.~Spolyar,
  ``Chain inflation via rapid tunneling in the landscape,''
  arXiv:hep-th/0612056.
  %%CITATION = HEP-TH/0612056;%%

%\cite{Huang:2007ek}
\bibitem{Huang:2007ek}
  Q.~G.~Huang,
  ``Simplified Chain Inflation,''
  JCAP {\bf 0705}, 009 (2007)
  [arXiv:0704.2835 [hep-th]].
  %%CITATION = JCAPA,0705,009;%%
  
  %\cite{Chialva:2008zw}
\bibitem{Chialva:2008zw}
  D.~Chialva and U.~H.~Danielsson,
  ``Chain inflation revisited,''
  JCAP {\bf 0810}, 012 (2008)
  [arXiv:0804.2846 [hep-th]].
  %%CITATION = JCAPA,0810,012;%%
     
%\cite{Ashoorioon:2008pj}
\bibitem{Ashoorioon:2008pj}
  A.~Ashoorioon, K.~Freese and J.~T.~Liu,
  ``Slow nucleation rates in Chain Inflation with QCD Axions or Monodromy,''
  arXiv:0810.0228 [hep-ph].
  %%CITATION = ARXIV:0810.0228;%%

%\cite{Ashoorioon:2008nh}
\bibitem{Ashoorioon:2008nh}
  A.~Ashoorioon and K.~Freese,
  ``Gravity Waves from Chain Inflation,''
  arXiv:0811.2401 [hep-th].
  %%CITATION = ARXIV:0811.2401;%%  %\cite{Khlebnikov:1997di}
  
%\cite{Komatsu:2008hk}
\bibitem{Komatsu:2008hk}
  E.~Komatsu {\it et al.}  [WMAP Collaboration],
  ``Five-Year Wilkinson Microwave Anisotropy Probe (WMAP)
  Observations:Cosmological Interpretation,''
  arXiv:0803.0547 [astro-ph].
  %%CITATION = ARXIV:0803.0547;%%
  
  %\cite{Watson:2006px}
\bibitem{Watson:2006px}
  S.~Watson, M.~J.~Perry, G.~L.~Kane and F.~C.~Adams,
  ``Inflation without inflaton(s),''
  JCAP {\bf 0711}, 017 (2007)
  [arXiv:hep-th/0610054].
  %%CITATION = JCAPA,0711,017;%%
  
   %\cite{Mukhanov:1990me}
\bibitem{Mukhanov:1990me}
  V.~F.~Mukhanov, H.~A.~Feldman and R.~H.~Brandenberger,
  ``Theory of cosmological perturbations. Part 1. Classical perturbations. Part
  2. Quantum theory of perturbations. Part 3. Extensions,''
  Phys.\ Rept.\  {\bf 215} (1992) 203.
  %%CITATION = PRPLC,215,203;%%

  %\cite{Bassett:2005xm}
\bibitem{Bassett:2005xm}
  B.~A.~Bassett, S.~Tsujikawa and D.~Wands,
  ``Inflation dynamics and reheating,''
  Rev.\ Mod.\ Phys.\  {\bf 78}, 537 (2006)
  [arXiv:astro-ph/0507632].
  %%CITATION = RMPHA,78,537;%%
  
  %\cite{Ohmori:2001am}
\bibitem{Ohmori:2001am}
  K.~Ohmori,
  ``A review on tachyon condensation in open string field theories,''
  arXiv:hep-th/0102085.
  %%CITATION = HEP-TH/0102085;%%
  
%\cite{Battefeld:2007en}
\bibitem{Battefeld:2007en}
  D.~Battefeld and T.~Battefeld,
  ``Non-Gaussianities in N-flation,''
  JCAP {\bf 0705}, 012 (2007)
  [arXiv:hep-th/0703012].
  %%CITATION = JCAPA,0705,012;%%   

%\cite{Kofman:2004yc}
\bibitem{Kofman:2004yc}
  L.~Kofman, A.~Linde, X.~Liu, A.~Maloney, L.~McAllister and E.~Silverstein,
  ``Beauty is attractive: Moduli trapping at enhanced symmetry points,''
  JHEP {\bf 0405}, 030 (2004)
  [arXiv:hep-th/0403001].
  %%CITATION = JHEPA,0405,030;%%

%\cite{Watson:2004aq}
\bibitem{Watson:2004aq}
  S.~Watson,
  ``Moduli stabilization with the string Higgs effect,''
  Phys.\ Rev.\  D {\bf 70}, 066005 (2004)
  [arXiv:hep-th/0404177].
  %%CITATION = PHRVA,D70,066005;%%
  
  
 
  %\cite{Gibbons:2006pa}
\bibitem{Gibbons:2006pa}
  G.~W.~Gibbons and N.~Turok,
  ``The measure problem in cosmology,''
  Phys.\ Rev.\  D {\bf 77}, 063516 (2008)
  [arXiv:hep-th/0609095].
  %%CITATION = PHRVA,D77,063516;%%

%\cite{Hartle:2007gi}
\bibitem{Hartle:2007gi}
  J.~B.~Hartle, S.~W.~Hawking and T.~Hertog,
  ``The No-Boundary Measure of the Universe,''
  Phys.\ Rev.\ Lett.\  {\bf 100}, 201301 (2008)
  [arXiv:0711.4630 [hep-th]].
  %%CITATION = PRLTA,100,201301;%%

%\cite{Linde:2007nm}
\bibitem{Linde:2007nm}
  A.~Linde,
  ``Towards a gauge invariant volume-weighted probability measure for eternal
  inflation,''
  JCAP {\bf 0706}, 017 (2007)
  [arXiv:0705.1160 [hep-th]].
  %%CITATION = JCAPA,0706,017;%%

 
  %\cite{Easther:2006vd}
\bibitem{Easther:2006vd}
  R.~Easther, J.~T.~.~Giblin and E.~A.~Lim,
  ``Gravitational Wave Production At The End Of Inflation,''
  Phys.\ Rev.\ Lett.\  {\bf 99}, 221301 (2007)
  [arXiv:astro-ph/0612294].
  %%CITATION = PRLTA,99,221301;%%
  
%%%%%%%%%%%%%%%%%%


%\cite{Chen:2006xjb}
\bibitem{Chen:2006xjb}
  X.~Chen, R.~Easther and E.~A.~Lim,
  ``Large non-Gaussianities in single field inflation,''
  JCAP {\bf 0706}, 023 (2007)
  [arXiv:astro-ph/0611645].
  %%CITATION = JCAPA,0706,023;%%

%\cite{Chen:2008wn}
\bibitem{Chen:2008wn}
  X.~Chen, R.~Easther and E.~A.~Lim,
  ``Generation and Characterization of Large Non-Gaussianities in Single Field
  Inflation,''
  JCAP {\bf 0804}, 010 (2008)
  [arXiv:0801.3295 [astro-ph]].
  %%CITATION = JCAPA,0804,010;%%
  
 %\cite{Linde:2007fr}
\bibitem{Linde:2007fr}
  A.~Linde,
  ``Inflationary Cosmology,''
  Lect.\ Notes Phys.\  {\bf 738}, 1 (2008)
  [arXiv:0705.0164 [hep-th]].
  %%CITATION = LNPHA,738,1;%%
  
  %\cite{Denef:2008wq}
\bibitem{Denef:2008wq}
  F.~Denef,
  ``Les Houches Lectures on Constructing String Vacua,''
  arXiv:0803.1194 [hep-th].
  %%CITATION = ARXIV:0803.1194;%%

  %\cite{Novello:2008ra}
\bibitem{Novello:2008ra}
  M.~Novello and S.~E.~P.~Bergliaffa,
  ``Bouncing Cosmologies,''
  Phys.\ Rept.\  {\bf 463}, 127 (2008)
  [arXiv:0802.1634 [astro-ph]].
  %%CITATION = PRPLC,463,127;%%

%\cite{Battefeld:2005av}
\bibitem{Battefeld:2005av}
  T.~Battefeld and S.~Watson,
  ``String gas cosmology,''
  Rev.\ Mod.\ Phys.\  {\bf 78}, 435 (2006)
  [arXiv:hep-th/0510022].
  %%CITATION = RMPHA,78,435;%%
  
  
%\cite{Brandenberger:2008nx}
\bibitem{Brandenberger:2008nx}
  R.~H.~Brandenberger,
  ``String Gas Cosmology,''
  arXiv:0808.0746 [hep-th].
  %%CITATION = ARXIV:0808.0746;%%
      
  %\cite{Lehners:2008vx}
\bibitem{Lehners:2008vx}
  J.~L.~Lehners,
  ``Ekpyrotic and Cyclic Cosmology,''
  Phys.\ Rept.\  {\bf 465}, 223 (2008)
  [arXiv:0806.1245 [astro-ph]].
  %%CITATION = PRPLC,465,223;%%
  
    
  %\cite{Hagedorn:1965st}
\bibitem{Hagedorn:1965st}
  R.~Hagedorn,
  ``Statistical thermodynamics of strong interactions at high-energies,''
  Nuovo Cim.\ Suppl.\  {\bf 3}, 147 (1965).
  %%CITATION = NUCUA,3,147;%%

  
%\cite{Allahverdi:2006iq}
\bibitem{Allahverdi:2006iq}
  R.~Allahverdi, K.~Enqvist, J.~Garcia-Bellido and A.~Mazumdar,
  ``Gauge invariant MSSM inflaton,''
  Phys.\ Rev.\ Lett.\  {\bf 97}, 191304 (2006)
  [arXiv:hep-ph/0605035].
  %%CITATION = PRLTA,97,191304;%%
  
  %\cite{Allahverdi:2006cx}
\bibitem{Allahverdi:2006cx}
  R.~Allahverdi, A.~Kusenko and A.~Mazumdar,
  ``A-term inflation and the smallness of the neutrino masses,''
  JCAP {\bf 0707}, 018 (2007)
  [arXiv:hep-ph/0608138].
  %%CITATION = JCAPA,0707,018;%%
  
  %\cite{Allahverdi:2006we}
\bibitem{Allahverdi:2006we}
  R.~Allahverdi, K.~Enqvist, J.~Garcia-Bellido, A.~Jokinen and A.~Mazumdar,
  ``MSSM flat direction inflation: slow roll, stability, fine tunning and
  %reheating,''
  JCAP {\bf 0706}, 019 (2007)
  [arXiv:hep-ph/0610134].
  %%CITATION = JCAPA,0706,019;%%

  %\cite{Allahverdi:2008bt}
\bibitem{Allahverdi:2008bt}
  R.~Allahverdi, B.~Dutta and A.~Mazumdar,
  ``Attraction towards an inflection point inflation,''
  Phys.\ Rev.\  D {\bf 78}, 063507 (2008)
  [arXiv:0806.4557 [hep-ph]].
  %%CITATION = PHRVA,D78,063507;%%
  
  %\cite{Jokinen:2004bp}
\bibitem{Jokinen:2004bp}
  A.~Jokinen and A.~Mazumdar,
  ``Inflation in large N limit of supersymmetric gauge theories,''
  Phys.\ Lett.\  B {\bf 597}, 222 (2004)
  [arXiv:hep-th/0406074].
  %%CITATION = PHLTA,B597,222;%%
  
  %\cite{Felder:1998vq}
\bibitem{Felder:1998vq}
  G.~N.~Felder, L.~Kofman and A.~D.~Linde,
  ``Instant preheating,''
  Phys.\ Rev.\  D {\bf 59}, 123523 (1999)
  [arXiv:hep-ph/9812289].
  %%CITATION = PHRVA,D59,123523;%%

%\cite{Enqvist:2003qc}
\bibitem{Enqvist:2003qc}
  K.~Enqvist, S.~Kasuya and A.~Mazumdar,
  ``MSSM Higgses as the source of reheating and all matter,''
  Phys.\ Rev.\ Lett.\  {\bf 93}, 061301 (2004)
  [arXiv:hep-ph/0311224].
  %%CITATION = PRLTA,93,061301;%%

  %%%%%%%%%%%%%%%%%%%%%%%%%%
%%%%%%%%%%%%%%%%%%%%%%%%%%%  

\bibitem{Khlebnikov:1997di}
  S.~Y.~Khlebnikov and I.~I.~Tkachev,
  ``Relic gravitational waves produced after preheating,''
  Phys.\ Rev.\  D {\bf 56}, 653 (1997)
  [arXiv:hep-ph/9701423].
  %%CITATION = PHRVA,D56,653;%%



 
  %\cite{Battefeld:2006sz}
\bibitem{Battefeld:2006sz}
  T.~Battefeld and R.~Easther,
  ``Non-gaussianities in multi-field inflation,''
  JCAP {\bf 0703}, 020 (2007)
  [arXiv:astro-ph/0610296].
  %%CITATION = JCAPA,0703,020;%%


  
%\cite{GarciaBellido:1998gm}
\bibitem{GarciaBellido:1998gm}
  J.~Garcia-Bellido,
  ``Preheating the universe in hybrid inflation,''
  arXiv:hep-ph/9804205.
  %%CITATION = HEP-PH/9804205;%%

%\cite{Easther:2006gt}
\bibitem{Easther:2006gt}
  R.~Easther and E.~A.~Lim,
  ``Stochastic gravitational wave production after inflation,''
  JCAP {\bf 0604}, 010 (2006)
  [arXiv:astro-ph/0601617].
  %%CITATION = JCAPA,0604,010;%%

%\cite{GarciaBellido:2007dg}
\bibitem{GarciaBellido:2007dg}
  J.~Garcia-Bellido and D.~G.~Figueroa,
  ``A stochastic background of gravitational waves from hybrid preheating,''
  Phys.\ Rev.\ Lett.\  {\bf 98}, 061302 (2007)
  [arXiv:astro-ph/0701014].
  %%CITATION = PRLTA,98,061302;%%

%\cite{GarciaBellido:2007af}
\bibitem{GarciaBellido:2007af}
  J.~Garcia-Bellido, D.~G.~Figueroa and A.~Sastre,
  ``A Gravitational Wave Background from Reheating after Hybrid Inflation,''
  Phys.\ Rev.\  D {\bf 77}, 043517 (2008)
  [arXiv:0707.0839 [hep-ph]].
  %%CITATION = PHRVA,D77,043517;%

%\cite{Dufaux:2007pt}
\bibitem{Dufaux:2007pt}
  J.~F.~Dufaux, A.~Bergman, G.~N.~Felder, L.~Kofman and J.~P.~Uzan,
  ``Theory and Numerics of Gravitational Waves from Preheating after
  Inflation,''
  Phys.\ Rev.\  D {\bf 76}, 123517 (2007)
  [arXiv:0707.0875 [astro-ph]].
  %%CITATION = PHRVA,D76,123517;%%

%\cite{Easther:2007vj}
\bibitem{Easther:2007vj}
  R.~Easther, J.~T.~Giblin and E.~A.~Lim,
  ``Gravitational Waves From the End of Inflation: Computational Strategies,''
  Phys.\ Rev.\  D {\bf 77}, 103519 (2008)
  [arXiv:0712.2991 [astro-ph]].
  %%CITATION = PHRVA,D77,103519;%%
  
 %\cite{Barnaby:2006cq}
\bibitem{Barnaby:2006cq}
  N.~Barnaby and J.~M.~Cline,
  ``Nongaussian and nonscale-invariant perturbations from tachyonic  preheating
  in hybrid inflation,''
  Phys.\ Rev.\  D {\bf 73} (2006) 106012
  [arXiv:astro-ph/0601481].
  %%CITATION = PHRVA,D73,106012;%%
  
  %\cite{Barnaby:2006km}
\bibitem{Barnaby:2006km}
  N.~Barnaby and J.~M.~Cline,
  ``Nongaussianity from tachyonic preheating in hybrid inflation,''
  Phys.\ Rev.\  D {\bf 75}, 086004 (2007)
  [arXiv:astro-ph/0611750].
  %%CITATION = PHRVA,D75,086004;%%

%\cite{Chambers:2008gu}
\bibitem{Chambers:2008gu}
  A.~Chambers and A.~Rajantie,
  ``Non-Gaussianity from massless preheating,''
  JCAP {\bf 0808}, 002 (2008)
  [arXiv:0805.4795 [astro-ph]].
  %%CITATION = JCAPA,0808,002;%%

%\cite{Enqvist:2004ey}
\bibitem{Enqvist:2004ey}
  K.~Enqvist, A.~Jokinen, A.~Mazumdar, T.~Multamaki and A.~Vaihkonen,
  ``Non-Gaussianity from Preheating,''
  Phys.\ Rev.\ Lett.\  {\bf 94}, 161301 (2005)
  [arXiv:astro-ph/0411394].
  %%CITATION = PRLTA,94,161301;%%
  
  %\cite{Enqvist:2005qu}
\bibitem{Enqvist:2005qu}
  K.~Enqvist, A.~Jokinen, A.~Mazumdar, T.~Multamaki and A.~Vaihkonen,
  ``Non-gaussianity from instant and tachyonic preheating,''
  JCAP {\bf 0503}, 010 (2005)
  [arXiv:hep-ph/0501076].
  %%CITATION = JCAPA,0503,010;%%

%\cite{Jokinen:2005by}
\bibitem{Jokinen:2005by}
  A.~Jokinen and A.~Mazumdar,
  ``Very Large Primordial Non-Gaussianity from multi-field: Application to
  Massless Preheating,''
  JCAP {\bf 0604}, 003 (2006)
  [arXiv:astro-ph/0512368].
  %%CITATION = JCAPA,0604,003;%%

%\cite{Silverstein:2003hf}
\bibitem{Silverstein:2003hf}
  E.~Silverstein and D.~Tong,
  ``Scalar Speed Limits and Cosmology: Acceleration from D-cceleration,''
  Phys.\ Rev.\  D {\bf 70}, 103505 (2004)
  [arXiv:hep-th/0310221].
  %%CITATION = PHRVA,D70,103505;%%
  
  %\cite{Alishahiha:2004eh}
\bibitem{Alishahiha:2004eh}
  M.~Alishahiha, E.~Silverstein and D.~Tong,
  ``DBI in the sky,''
  Phys.\ Rev.\  D {\bf 70}, 123505 (2004)
  [arXiv:hep-th/0404084].
  %%CITATION = PHRVA,D70,123505;%%

%\cite{Chen:2004gc}
\bibitem{Chen:2004gc}
  X.~Chen,
  ``Multi-throat brane inflation,''
  Phys.\ Rev.\  D {\bf 71}, 063506 (2005)
  [arXiv:hep-th/0408084].
  %%CITATION = PHRVA,D71,063506;%%

%\cite{Chen:2005ad}
\bibitem{Chen:2005ad}
  X.~Chen,
  ``Inflation from warped space,''
  JHEP {\bf 0508}, 045 (2005)
  [arXiv:hep-th/0501184].
  %%CITATION = JHEPA,0508,045;%%

%\cite{Easson:2007dh}
\bibitem{Easson:2007dh}
  D.~A.~Easson, R.~Gregory, D.~F.~Mota, G.~Tasinato and I.~Zavala,
  ``Spinflation,''
  JCAP {\bf 0802}, 010 (2008)
  [arXiv:0709.2666 [hep-th]].
  %%CITATION = JCAPA,0802,010;%%

%\cite{Langlois:2008wt}
\bibitem{Langlois:2008wt}
  D.~Langlois, S.~Renaux-Petel, D.~A.~Steer and T.~Tanaka,
  ``Primordial fluctuations and non-Gaussianities in multi-field DBI
  inflation,''
  Phys.\ Rev.\ Lett.\  {\bf 101}, 061301 (2008)
  [arXiv:0804.3139 [hep-th]].
  %%CITATION = PRLTA,101,061301;%%

%\cite{Langlois:2008qf}
\bibitem{Langlois:2008qf}
  D.~Langlois, S.~Renaux-Petel, D.~A.~Steer and T.~Tanaka,
  ``Primordial perturbations and non-Gaussianities in DBI and general
  multi-field inflation,''
  arXiv:0806.0336 [hep-th].
  %%CITATION = ARXIV:0806.0336;%%
  
  %\cite{ArmendarizPicon:1999rj}
\bibitem{ArmendarizPicon:1999rj}
  C.~Armendariz-Picon, T.~Damour and V.~F.~Mukhanov,
  ``k-Inflation,''
  Phys.\ Lett.\  B {\bf 458}, 209 (1999)
  [arXiv:hep-th/9904075].
  %%CITATION = PHLTA,B458,209;%%

%\cite{Garriga:1999vw}
\bibitem{Garriga:1999vw}
  J.~Garriga and V.~F.~Mukhanov,
  ``Perturbations in k-inflation,''
  Phys.\ Lett.\  B {\bf 458}, 219 (1999)
  [arXiv:hep-th/9904176].
  %%CITATION = PHLTA,B458,219;%%




  \end{thebibliography}
\end{document}